\documentclass[5p,lefttitle]{elsarticle}

\usepackage{graphicx}
\usepackage{tabularx}
\usepackage{booktabs}
\usepackage{array}
\usepackage{dsfont}
\usepackage{amsmath,amssymb}
\usepackage{array}
\usepackage{longtable}
\usepackage{calc}
\usepackage{float}
\usepackage{needspace}
\usepackage{multirow}
\usepackage{url}
\usepackage{hyperref}

\usepackage{listings}
\providecommand{\pkg}[1]{\textsf{#1}}

\title{Using the \texttt{spsurv} \textsf{R}~package for semi-parametric time-to-event analysis}

\author[inst1]{Renato Valladares Panaro\corref{cor1}}
\ead{rvpanaro@gmail.com}

\author[inst2]{Vin\'{i}cius Mayrink}
\ead{vdinizm@gmail.com}

\author[inst2]{F\'{a}bio Demarqui}
\ead{fndemarqui@gmail.com}

\cortext[cor1]{Corresponding author}

\address[inst1]{Department of Medical Statistics, University Medical Center G\"{o}ttingen, Georg-August-Universit\"{a}t G\"{o}ttingen, 37073 G\"{o}ttingen, Germany}
\address[inst2]{Departamento de Estat\'{i}stica, Instituto de Ci\^{e}ncias Exatas, Universidade Federal de Minas Gerais, Av.\ Ant\^{o}nio Carlos 6627, 31270-901 Belo Horizonte, MG, Brazil}
\date{\today}

\begin{document}
\begin{abstract}
We present \pkg{spsurv}, an \textsf{R} package for semi-parametric time-to-event regression based on Bernstein-polynomial estimation of unknown baseline functions. The package provides a unified modelling interface for proportional hazards (PH), proportional odds (PO), and accelerated failure time (AFT) models for right-censored data, with either maximum likelihood or Bayesian estimation via Stan. Smooth baseline hazard, odds-function, or log-time structures are estimated without assuming a parametric baseline family, while retaining familiar hazard-ratio, odds-ratio, and time-ratio interpretations. We describe methodology, implementation, and syntax; evaluate finite-sample behaviour in a Monte Carlo study; and illustrate usage with oncology trials.
\end{abstract}
\maketitle


\setcounter{section}{0}
\renewcommand{\thesection}{\arabic{section}}
\newcounter{none}

\section{Introduction: Semi-parametric survival analysis in R}\label{sec:intro}

Several R packages \citep{R:2025} support survival regression, each built around a different modelling goal. \pkg{survival} \citep{survival:2000} is the standard tool for Cox regression and classical parametric/AFT models; \pkg{flexsurv} \citep{flexsurv:2016} emphasises flexible parametric survival models; \pkg{timereg} \citep{timereg} offers semi-parametric regression under alternative estimating frameworks; \pkg{ggsurvfit} \citep{ggsurvfit:2024} provides time-to-event visualisation tools that complement any of these; and \pkg{spBayesSurv} \citep{Zhou:2021} implements Bayesian semi-parametric survival models whose baseline is centred on a parametric family. \pkg{spsurv} \citep{spsurv:2026} occupies a different niche: it provides one interface for PH, PO, and AFT models under both maximum likelihood and Bayesian estimation. This unified design is motivated by two practical facts, the proportional hazards assumption is frequently violated in applied work \citep{kalbfleisch2002statistical}, and choosing between regression families is easier when every candidate model is fit on the same data within a consistent framework. Estimation in \pkg{spsurv} uses full likelihoods with smooth Bernstein-polynomial (BP) baselines for the hazard, odds, or log-time structure.

The BP itself is well established for regression \citep{Tenbusch:1997, Chang:2007} and density estimation \citep{Vitale:1975, Petrone:1999, Babu:2002}, but its application to survival analysis specifically has received comparatively less attention, and almost always as an alternative to Cox's partial-likelihood estimator \citep{Cox:1972}. \citet{Chang:2005} treated the polynomial degree as random and modelled the failure rate with a Beta-process prior for homogeneous populations. \citet{Osman:2012} used sieve estimation based on Bernstein approximation to recover the baseline hazard, and \citet{Mclain:2013} applied the same idea to time-transformation models. \citet{Chen:2014} centred a transformed BP at a standard parametric family to build a Bayesian nonparametric prior for smooth densities in the accelerated hazards model, using a Dirichlet process because the polynomial degree itself was treated as random. \citet{Zhou:2018} placed a prior directly on the baseline hazard function. The \pkg{spBayesSurv} package \citep{Zhou:2021} follows a related strategy, assuming a parametric distribution (e.g., Weibull or log-logistic) as the central model and using the BP to allow flexible deviations around it. \pkg{spsurv} differs from each of these in scope: rather than committing to one baseline-modelling strategy for one regression family, it applies the same BP machinery across PH, PO, and AFT alike.

We describe methodology and implementation, report a Monte Carlo check of
coefficient recovery under standard aligned generators, and give two clinical
illustrations. A fuller study of baseline shape, polynomial degree, and
non-positive-definite Hessians is left to future work. Three features distinguish \pkg{spsurv} from related
software. First, PH, PO, and AFT models share a common fitting interface, allowing analysts to compare model families without changing data preparation. Second, covariates are standardised internally to improve numerical conditioning and results are returned on the original covariate scale (Section~\ref{sec:varscale}). Third, both maximum-likelihood and Bayesian estimation are available within the same framework, with uncertainty handled separately for the two approaches (Sections~\ref{sec:delta-method}--\ref{sec:priors}). The computational differences between the PH/PO and AFT implementations are described in Section~\ref{sec:bp-domain}.

The remainder of the paper is organised as follows. Section
\ref{sec:models} introduces notation for right-censored data and
describes the Bernstein-based likelihood formulations for PH, PO, and
AFT models, including implementation details. Section
\ref{sec:simulation} reports a Monte Carlo study of Bernstein-based
fits across aligned generator--model settings. Section
\ref{sec:illustrations} demonstrates the package on two clinical data
sets. Section \ref{sec:discussion} concludes the study.

\section{Models and software}\label{sec:models}

This section introduces the BP framework used to model baseline
functions in semi-parametric survival regression. The key idea is to
approximate the unknown baseline function with a finite-degree
polynomial. We present key time-to-event analysis identities by linking
cumulative-function representations to other quantities such as hazard,
odds, and log survival times, through derivative properties. The
resulting sieve gives a smooth baseline without specifying a parametric
family: higher degrees can increase flexibility. For survival models,
the BP derivative property is useful because many baseline quantities
are cumulative (e.g., cumulative hazards \(H_0\) or odds function \(R_0 = F_0(t)/S_0(t)\)) and must be monotone by
construction, which non-negative BP parameters guarantee when estimating
the baseline from the data.

\subsection{Regression model classes: PH, PO, and AFT}

The \texttt{spsurv} package implements three common classes of survival
regression models: proportional hazards (PH), proportional odds (PO), and
accelerated failure time (AFT). These models all relate the time until an
event to a set of covariates, but they differ in the scale on which the
covariates act. General introductions to these model classes can be found in
standard survival-analysis texts \citep{Klein:1997,kalbfleisch2002statistical,Collett2015}.

In the proportional hazards model, covariates act on the instantaneous event
rate. The hazard function \(h(t \mid x)\) represents the event rate at time
\(t\), among subjects who have survived up to that time. If \(h_0(t)\) is the
baseline hazard, the PH model is written as
{\begin{align}
h(t \mid x) = h_0(t)e^{\beta'x}.
\end{align}\normalsize}
Thus, \(e^{\beta_j}\) is interpreted as a hazard ratio. A value larger than
one indicates a higher instantaneous risk of the event for a one-unit increase
in the covariate, holding the other covariates fixed; a value smaller than one
indicates a lower instantaneous risk. The term proportional hazards means that
this hazard ratio is constant over time, although the baseline hazard
\(h_0(t)\) itself may vary with time \citep{Cox:1972}.

In the proportional odds model, covariates act on the odds function of
failure rather than on the instantaneous hazard. Let \(R_0(t)\) denote the
baseline odds function. The PO model can be written as
{\begin{align}
S(t \mid x) = \{1 + R_0(t)e^{\beta'x}\}^{-1}.
\end{align}\normalsize}
In this model, \(e^{\beta_j}\) is interpreted as an odds ratio for the
odds function of having experienced the event by time \(t\). Therefore, the
PO model imposes proportionality on the odds-function scale rather than on
the hazard scale \citep{Bennett:1983}. This provides an alternative regression
structure when the PH interpretation or assumption is not appropriate.

In the accelerated failure time model, covariates act directly on the
survival-time scale rather than on instantaneous risk or odds function of
failure: each covariate accelerates or decelerates the time to the event, and
\(e^{\beta_j}\) is interpreted as a time ratio. In \pkg{spsurv}, this structure
is written in terms of log-time residuals,
{\begin{align}
w_i = \log(y_i) - \beta'x_i,
\end{align}\normalsize}
where \(y_i\) is the observed follow-up time and \(x_i\) is the covariate
vector for subject \(i\). The baseline component is then defined through the
distribution of \(w_i\). A value of this ratio larger than one indicates that
the covariate is associated with longer survival times, whereas a value
smaller than one indicates shorter survival times
\citep{Prentice:1973,kalbfleisch2002statistical}.

Therefore, PH, PO, and AFT models answer related but distinct scientific
questions. PH models compare instantaneous event rates, PO models compare
odds function of failure, and AFT models compare survival times. The
Bernstein-polynomial versions implemented in \texttt{spsurv} preserve these
standard interpretations while replacing the unknown baseline hazard,
odds function, or log-time residual distribution with a flexible polynomial
approximation.

\subsection{Bernstein-polynomials in time-to-event analysis}\label{bernstein-polynomials-in-time-to-event-analysis}

In the time-to-event regression models considered here, the same
construction is used to represent the baseline function. The derivative
property is central and allows the derivative of a Bernstein
approximation to be expressed as a weighted sum of beta densities. This
representation lets us model baseline hazards and cumulative hazards
within a survival-regression framework; see also \citet{Osman:2012}. Let \(c:=H_0\) be the cumulative baseline hazard on
\([\tau_a,\tau_b]=[0,\tau]\). The derivative of its Bernstein
approximation can be written as

{\footnotesize\begin{align}
dB_m^{H_0}(t)
&=
\frac{m}{\tau}
\sum_{i=0}^{m-1}
\left\{
H_0\!\left(\frac{i+1}{m}\tau\right)
-
H_0\!\left(\frac{i}{m}\tau\right)
\right\}
b_{i,m-1}(t/\tau)\,dt \nonumber \\
&=
\sum_{k=1}^m
\left\{
H_0\!\left(\frac{k}{m}\tau\right)
-
H_0\!\left(\frac{k-1}{m}\tau\right)
\right\}
\frac{f_{\beta}(t/\tau\mid k,\;m-k+1)}{\tau}\,dt \nonumber \\
&=
\sum_{k=1}^m \gamma_k g_{k,m}(t) =\boldsymbol{\gamma}'\boldsymbol{g}_{m}(t)\,dt,
\end{align}\normalsize\label{eq:deriv} where
\(g_{k,m}(t)=f_{\beta}(t/\tau\mid k,\;m-k+1)\tau^{-1},\) and
\(f_{\beta}(\cdot\mid a,b)\) denotes the beta density with shape
parameters \(a\) and \(b\). The Bernstein coefficients are positive differences between cumulative hazard functions given by
\(\gamma_k=H_0\!\left(\frac{k}{m}\tau\right)-H_0\!\left(\frac{k-1}{m}\tau\right), k=1,\dots,m.\)
These coefficients depend on the polynomial degree \(m\) but not on time
\(t\). In practice, because the true cumulative hazard is unknown, the
coefficients \(\gamma_1,\dots,\gamma_m\) are estimated from the data for a
chosen degree \(m\).

To incorporate censored observations, let \((Y_1,\dots,Y_n)'\) denote the
vector of observed follow-up times with \(Y_i=\min(T_i,C_i)\), where \(T_i\)
is the event time, \(C_i\) is the censoring time, and
\(\delta_i=I(T_i\le C_i)\) is the event indicator. Let
\(S(y\mid\boldsymbol\theta)=P(T> y\mid\boldsymbol\theta)\) and
\(h(y\mid\boldsymbol\theta)\) denote the event-time survival and hazard
functions, respectively, where \(\boldsymbol\theta\) generically collects the
baseline-function parameters in Table \ref{tab:models} (either \(\boldsymbol{\gamma}\) or \(\boldsymbol{\xi}\)) and \(\boldsymbol\beta\) the regression
coefficients. Under non-informative censoring, the full likelihood is
given by
{\begin{align}
p(\boldsymbol{y} \vert \boldsymbol \theta, \boldsymbol{\beta})
&\propto \prod\limits_{i = 1}^{n} h(y_i \vert \boldsymbol \theta,\boldsymbol{\beta})^{\delta_i}  S(y_i \vert \boldsymbol \theta,\boldsymbol{\beta}),
\end{align}\normalsize}
where \(\boldsymbol{y}=(y_1, y_2, \dots, y_n)'\) is the
vector of observed times and \(\boldsymbol \beta\) is the vector of
regression coefficients, see \citet{Klein:1997}. Table~\ref{tab:models}
collects the resulting baseline approximation and survival/hazard
expressions for all three model classes side by side, and is the
reference point for the implementation details in
Sections~\ref{sec:bp-domain}--\ref{sec:delta-method}.

\begin{table}[!h]
\footnotesize
\centering
\caption{Likelihood structures in \texttt{spsurv}. A degree-$m$ Bernstein basis $\boldsymbol{g}_m(\cdot)$ (cumulative $\boldsymbol{G}_m(\cdot)$, coefficients $\boldsymbol{\gamma}$ or $\boldsymbol{\xi}$) approximates the baseline quantities listed per model. BPPH: $h_0(y)$ and $H_0(y)$; BPPO: $R_0(y)$ and $r_0(y)$; BPAFT: $h_0^W(w)$ and $H_0^W(w)$ for $w_i=\log(y_i)-\boldsymbol{\beta}'\boldsymbol{x}_i$, with observed-time hazard $y_i^{-1}h_0^W(w_i)$.}
\begin{tabularx}{\columnwidth}{%
>{\raggedright\arraybackslash}p{0.1\columnwidth}
>{\raggedright\arraybackslash}p{0.325\columnwidth}
>{\raggedright\arraybackslash}X}
\toprule
Model & Baseline function approximated & Survival / hazard structure \\
\midrule
BPPH
&
\(h_0(y)=\boldsymbol{\gamma}'\boldsymbol g_m(y)\),
\(H_0(y)=\boldsymbol{\gamma}'\boldsymbol G_m(y)\)
&
\(h(y\vert\boldsymbol x)=h_0(y)e^{\boldsymbol\beta'\boldsymbol x}\);
\(S(y\vert\boldsymbol x)=e^{-H_0(y)\exp(\boldsymbol\beta'\boldsymbol x)}\)
\\[0.8em]
& & \\
BPPO
&
\(R_0(y)=\boldsymbol{\xi}'\boldsymbol G_m(y)\),
\(r_0(y)=\boldsymbol{\xi}'\boldsymbol g_m(y)\)
&
\(S(y\vert\boldsymbol x)=\{1+R_0(y)e^{\boldsymbol\beta'\boldsymbol x}\}^{-1}\)
\\[0.8em]
& & \\
BPAFT
&
Log-time residuals baseline via
\(h^{W}_0(w)=\boldsymbol{\gamma}'\boldsymbol g_m(w)\),
\(H^{W}_0(w)=\boldsymbol{\gamma}'\boldsymbol G_m(w)\)
&
\(w_i=\log(y_i)-\boldsymbol\beta'\boldsymbol x_i\);
\(S(y_i\vert\boldsymbol x_i)=S_0^{W}(w_i)\);
\(h(y_i\vert\boldsymbol x_i)=y_i^{-1}h_0^{W}(w_i)\)
\\
\bottomrule
\end{tabularx}\label{tab:models}
\end{table}

\subsubsection{Time rescaling and an efficient polynomial representation}\label{sec:bp-domain}

BPs are originally defined on the unit interval. Therefore, when they
are used to model functions of time or log-time, the relevant domain
must first be transformed. For ordinary time scales, this is
straightforward: the observed time range can be bounded by
\(\tau=\max(y_i)\), giving the interval \([0,\tau]\), which is then mapped
linearly to \([0,1]\). This is the approach used in the BPPO and BPPH
models, where the time axis is fixed independently of the parameters.
The function \texttt{bp.basis()} computes the polynomial basis matrices
that, together with the vector of Bernstein coefficients, represent the
baseline component of the model. More specifically, \texttt{bp.basis()}
returns the hazard and cumulative-hazard basis matrices
\(\boldsymbol{g}_m\) and \(\boldsymbol{G}_m\), built from scaled beta
densities and beta cumulative distribution functions evaluated at
\(y/\tau\).

The key difference for AFT models is that the time scale itself depends on the regression coefficients. As \(\beta\) changes during estimation, the log-time residuals change, so the interval used to construct the Bernstein basis must also be updated. This makes BPAFT computationally more involved than BPPH and BPPO.
Specifically, the AFT baseline is evaluated on the log-time residuals
{\begin{align}
w_i = \log(y_i)-\beta'x_i .
\end{align}\normalsize}
Because these residuals depend on the regression coefficients, the interval
used to map them to the unit domain also depends on the current value of
\(\beta\). At a given parameter value, let
{\begin{align}
w_{\min}(\beta)=\min_i\{\log(y_i)-\beta'x_i\},\nonumber
\\
w_{\max}(\beta)=\max_i\{\log(y_i)-\beta'x_i\}.
\end{align}\normalsize}
The residuals are then mapped to the unit interval through the affine
transformation
{\begin{align}
u_i(\beta)
=
\frac{w_i-w_{\min}(\beta)}
     {w_{\max}(\beta)-w_{\min}(\beta)} ,
\qquad 0\leq u_i(\beta)\leq 1 .
\end{align}\normalsize}
Thus, the polynomial baseline is evaluated at \(u_i(\beta)\), but the
mapping itself must be recomputed whenever \(\beta\) changes. This differs
from the PH and PO models, where the time range \([0,\tau]\), with
\(\tau=\max_i y_i\), is fixed by the observed data and does not depend on the
regression coefficients.

Full expressions, together with a covariate-scaled-time formulation \citep{kalbfleisch2002statistical}, are given
in~\ref{app:aft-gradients}. For each new value of \(\boldsymbol{\beta}\), the
endpoints \(w_{\min}(\boldsymbol{\beta})\) and
\(w_{\max}(\boldsymbol{\beta})\), the scaled arguments
\(u_i(\boldsymbol{\beta})\), and hence the entire basis must be
re-evaluated. Because the scaled residuals change with \(\beta\), BPAFT cannot use the
precomputed beta-mixture basis available for BPPH and BPPO. Instead, each
beta density \(g_{k,m}\) on the unit interval is written as a monomial
expansion in \(u\). The transformation matrix depends only on the
polynomial degree and is computed once (\texttt{pw.basis()}); only the
powers \(u^{0},\ldots,u^{m-1}\) must be updated as \(\beta\) changes. This
avoids repeated beta-density evaluations during estimation. The algebraic
map is given in~\ref{app:monomial}.

\subsection{Implementation details} \label{sec:implementation}

\subsubsection{Covariate centring and standardisation}\label{sec:varscale}

The estimation is carried out using standardised
covariates rather than the original covariates (\texttt{scale = TRUE} in
\texttt{spbp.default()}). Specifically, if
\(\boldsymbol{x}_i\) denotes the original covariate vector for subject
\(i\), the standardised covariates are defined as \(\boldsymbol{z}_i =
\boldsymbol{s}_x^{-1} \circ
(\boldsymbol{x}_i - \overline{\boldsymbol{x}}),\) where
\(\overline{\boldsymbol{x}}\) is the vector of sample means,
\(\boldsymbol{s}_x\) is the vector of sample standard deviations, and
\(\circ\) denotes the Hadamard (element-wise) product. The models are
therefore fitted in terms of a standardised parameterisation
\((\boldsymbol{\psi}, \boldsymbol{\eta})\), where \(\boldsymbol{\psi}\)
contains the baseline Bernstein parameters and \(\boldsymbol{\eta}\) the
regression coefficients associated with the standardised covariates.
After estimation, these quantities are transformed back to the original
scale to recover the parameters of scientific interest.

In all three models, the regression coefficients are re-scaled to the
original covariate scale according to
\(\boldsymbol{\beta} = \boldsymbol{\eta} \circ \boldsymbol{s}_x^{-1}.\)
For the BPPH and BPPO models, centring the
covariates also shifts the baseline component, so the Bernstein coefficients must be adjusted accordingly. The
corresponding back-transformation is given by

{
\begin{align}
g &\colon \mathbb{R}_+^m \times \mathbb{R}^p
\to \mathbb{R}_+^m \times \mathbb{R}^p \nonumber \\
(\boldsymbol{\psi}, \boldsymbol{\eta})'
&\mapsto
(\boldsymbol{\gamma}, \boldsymbol{\beta})'
=
\begin{cases}\footnotesize
\left(
\boldsymbol{\psi}e^{-\boldsymbol{\eta}'(\boldsymbol{s}_x^{-1} \circ
\overline{\boldsymbol{x}})},
\ \boldsymbol{\eta} \circ \boldsymbol{s}_x^{-1}
\right)', \text{for PH, PO}, \\[1ex]
\footnotesize\left(
\boldsymbol{\psi},
\ \boldsymbol{\eta} \circ \boldsymbol{s}_x^{-1}
\right)',   \text{for AFT}.\label{transf}
\end{cases}
\end{align}\normalsize}
Covariate standardisation affects the regression coefficients in all
models, but affects the baseline Bernstein parameters only in the BPPH
and BPPO models. In the BPAFT model, those baseline parameters remain
unchanged. The next two subsections use this standardised
parameterisation in two different ways: Section~\ref{sec:delta-method} maps the
observed information matrix back through \(g(\cdot)\) for
maximum-likelihood inference, and Section~\ref{sec:priors} places priors
directly on \((\boldsymbol{\psi}, \boldsymbol{\eta})'\) for Bayesian
estimation.

\subsubsection{Delta-method variance estimation}\label{sec:delta-method}

Maximum-likelihood estimation is performed on the internally
standardised parameterisation described in Section~\ref{sec:varscale},
whereas inference is reported on the original covariate scale.
Likelihood-ratio statistics are invariant to this transformation, but
Wald standard errors and confidence intervals depend on the
parameterisation. We therefore use the multivariate delta method to
transform the estimated covariance matrix back to the original scale.

Maximum-likelihood estimates are obtained using the
\texttt{rstan::optimizing()} call. After
fitting, the covariance matrix is transformed through the
back-transformation \(g(\cdot)\) in \eqref{transf}, and the delta method gives the approximate covariance matrix.
The observed Hessian is inverted block-wise, using separate blocks for the
regression coefficients \(\beta\) and the basis parameters \(\gamma\); this
is a numerical stabilisation strategy for cases in which the full observed
information matrix is singular or nearly singular, so that direct inversion
may be unreliable \citep{nocedal2006numerical}. Further detail on the
conditioning strategy for each block is given in~\ref{app:hessian-blocks}.

For likelihood estimation of survival curves, the cumulative hazard
\(H(t\mid\boldsymbol{x})\) (with the PH/PO/AFT constructions in
Table~\ref{tab:models}) is treated as a smooth function of
\((\boldsymbol{\beta},\boldsymbol{\gamma})\). Pointwise variance is obtained
from the joint covariance after back-transformation, using gradients with
respect to both blocks. Under PH and PO, covariate centring implies
cross-derivatives \(\partial \gamma_k / \partial \eta_j =
-\gamma_k\,\bar{x}_j/s_j\) in the Jacobian of \(g\), where \(\eta_j\) is the
standardised coefficient, and \(\bar{x}_j\) and
\(s_j\) are the sample mean and standard deviation of covariate \(j\).
For BPAFT, \(\partial H/\partial\boldsymbol{\beta}\) follows the endpoint-corrected
gradient of~\ref{app:aft-gradients}, evaluated at the MLE training-set
endpoints \(i^\ast\) and \(j^\ast\); omitting the endpoint correction can
materially inflate delta-method standard errors on moderate-sized data sets
even when the \(\boldsymbol{\gamma}\) information block is well conditioned.
Survival standard errors are computed on the log scale and passed to
\texttt{survfit.spbp}-style intervals (default \texttt{type = "log"}).

In the software, \texttt{survfit()} and \texttt{predict()} issue a
warning when this instability is detected. Regression-coefficient
intervals remain available when the \(\beta\) block is well identified,
whereas \texttt{vcov(..., bp.param = TRUE)} returns \texttt{NA} for
unstable Bernstein-parameter entries. The Bayesian \texttt{survfit.spbp}
does not use the delta method; credible bands are obtained directly
from posterior survival curves.

\subsubsection{Weakly informative default priors}\label{sec:priors}

For Bayesian estimation, internal standardisation of the covariates is
useful because it puts coefficients on a common scale, so that each
coefficient represents the effect of a one-standard-deviation increase
in the corresponding covariate scale \citep{gelman2008weakly}. We write
\(\mathrm{Normal}(0,\sigma)\) for a normal distribution with variance
\(\sigma^2\), so the standard deviation is \(\sigma\). Under this
parameterisation, \(\eta_j \sim \mathrm{Normal}(0,2)\) places
approximately 95\% of its prior mass between \(-4\) and \(4\). On the
exponentiated scale, this corresponds to ratios of approximately \(0.018\)
to \(54.6\). The default is therefore weakly informative: it allows large
effects while providing some regularisation against extreme coefficient
values. In the
Stan implementation, the default prior for standardised regression
coefficients is passed as \texttt{normal(0, 2)}, because Stan's second
\texttt{normal} parameter is the standard deviation.
A weaker prior for \(\eta_j\), such as \(\eta_j \sim \mathrm{Normal}(0,4)\)
or \(\eta_j \sim \mathrm{Normal}(0,10)\), is possible, but it allows
very large values on the hazard, odds, or time-ratio scale and can
reduce sampling efficiency.

Standardisation of the covariates also affects the scale of the
Bernstein coefficients in the PH and PO models, which means a poorly
chosen weakly informative prior can make baseline estimation harder: a
prior that discourages highly variable baseline estimates, that is,
very large BP parameters \(\boldsymbol{\gamma}\), needs to account for
the fact that values of those basis parameters close to zero are often
the most plausible, so their corresponding logarithms are typically
negative, sometimes substantially so. \pkg{spsurv} therefore uses
\(\log(\psi_k) \sim \mathrm{Normal}(0,4)\) as a weakly informative prior on the
baseline scale, alongside the default regression prior
\(\eta_j \sim \mathrm{Normal}(0,2)\) on the standardised scale. Tightening the
regression prior, for example to \(\eta_j \sim \mathrm{Normal}(0,1)\),
reduces posterior standard deviations, but at the cost of a greater risk
of prior--data conflict.

In practice, sensitivity to the prior scale can be assessed by comparing
results under alternative weakly informative choices, for example
\(\mathrm{Normal}(0,2)\) versus \(\mathrm{Normal}(0,4)\) or
\(\mathrm{Normal}(0,10)\). In moderate samples, posterior coefficient
summaries often change little across such priors, whereas looser priors
can increase Monte Carlo variability; the default \(\mathrm{Normal}(0, 2)\)
specification was chosen to allow broad effects while avoiding the most
extreme prior scales considered here.

\section{Monte Carlo simulation}\label{sec:simulation}

\subsection{Goal and limitations}\label{goal-and-limitations}

The goal is to check whether the package recovers regression coefficients
when data are fitted with the same regression family used to generate them, which is the
intended routine use of \pkg{spsurv}. The primary demonstration uses
Weibull generators, whose hazards are monotone at the chosen shape.
Log-logistic cells are included for completeness. Log-logistic belongs
to the PO and AFT families, not PH \citep{Collett2015}; with shape \(1.5\)
its hazard is unimodal. Fitting BPPH to log-logistic PH data is therefore
a stress case for an uncentred Bernstein sieve, not a claim that a
parametric log-logistic PH model exists. The design does not assess
baseline-shape recovery, link misspecification, Hessian repair, or
data-adaptive degree rules, which remain open for future work.
Table \ref{tab:mc-design-events} reports the simulation design and
realised event percentages. Failure times were generated with the
\texttt{rsurv} package \citep{rsurv} using three generator functions:
\texttt{rphreg} for PH data, \texttt{rporeg} for PO data, and
\texttt{raftreg} for AFT data. In each case, the fitted model used the
same regression link as the generator.

Fits use \(m=n^{0.4}\), slightly below the package default
\(m=\sqrt{n}\), so the main table is conservative relative to default
use. The inferential target is the regression coefficients, summarised
by relative bias (\%), coverage, and SE calibration. Direct
baseline-function recovery error is not reported.

Two baseline families were considered: Weibull and log-logistic. Thus, the six generator cells were Weibull-PH, Weibull-PO, Weibull-AFT, log-logistic-PH, log-logistic-PO, and log-logistic-AFT. Covariates were generated independently in every replicate, with
\(x_{i1} \sim \mathrm{Normal}(0,1)\) and
\(x_{i2} \sim \mathrm{Bernoulli}(0.5)\), for \(i=1,\ldots,n\). The same covariate-generation mechanism was used for all generator families, model classes, sample sizes, and estimation approaches. The data-generating regression coefficients were likewise held fixed at
$\beta_0 = (\beta_{01}, \beta_{02})' = (-2, 1)'$ across all generator cells,
where $\beta_{01}$ is the coefficient of the continuous covariate (labelled
\emph{age} in Table~\ref{tab:bp-mcsim-abc}) and $\beta_{02}$ that of the binary
covariate (labelled \emph{sex}). These are the values of $\beta_0$ entering the relative-bias definition in~\eqref{eqn:relbias}. Independent administrative censoring was generated as
\(C_i \sim \mathrm{Uniform}(0,10)\), and the observed data were \(Y_i=\min(T_i,C_i)\) and \(\delta_i=I(T_i\leq C_i)\).
The same censoring mechanism was used for all generator families and model classes.

\begin{table}[!h]
\footnotesize
\centering
\setlength{\tabcolsep}{3pt}
\renewcommand{\arraystretch}{1.05}
\caption{\label{tab:mc-design-events} Monte Carlo design for Table~\ref{tab:bp-mcsim-abc}. Mean event percentages (\(100 \times \mathrm{E}(\#\text{events}/n)\)); MLE and Bayes share the same realisation; mean censoring is approximately \(100 -\) event rate.}
\begin{tabularx}{\columnwidth}{@{}
  >{\raggedright\arraybackslash}p{0.16\columnwidth}
  >{\raggedright\arraybackslash}p{0.07\columnwidth}
  >{\raggedleft\arraybackslash}p{0.07\columnwidth}
  >{\raggedleft\arraybackslash}p{0.07\columnwidth}
  >{\raggedleft\arraybackslash}p{0.11\columnwidth}
  >{\raggedleft\arraybackslash}p{0.11\columnwidth}
  >{\raggedleft\arraybackslash}p{0.11\columnwidth}
  >{\raggedleft\arraybackslash}p{0.11\columnwidth}
@{}}
\toprule
Family &
Class &
Shape &
Scale &
Event \% (\(n=50\)) &
Event \% (\(n=100\)) &
Event \% (\(n=200\)) &
Event \% (\(n=500\)) \\
\midrule
Weibull      & PH  & 1.5 & 1 & 86.3 & 86.2 & 86.4 & 86.4 \\
Weibull      & PO  & 1.5 & 1 & 91.3 & 91.3 & 91.3 & 91.3 \\
Weibull      & AFT  & 1.5 & 1 & 68.8 & 69.3 & 69.4 & 69.3 \\
Log-logistic      & PH  & 1.5 & 1 & 74.8 & 74.5 & 74.6 & 74.6 \\
Log-logistic      & PO  & 1.5 & 1 & 79.6 & 79.2 & 79.5 & 79.5 \\
Log-logistic      & AFT  & 1.5 & 1 & 62.6 & 62.9 & 63.1 & 63.0 \\
\bottomrule
\end{tabularx}
\end{table}

For each generator cell, we used \(R=1000\) Monte Carlo replicates at sample sizes \(n\in\{50,100,200,500\}\). Both maximum likelihood and Bayesian fits were applied to the same simulated data within each replicate. Bayesian summaries use all 1000 replicates; MLE summaries omit replicates with non-finite estimates or intervals, which is most visible in the AFT cells. The Bernstein-polynomial degree was set to \(m= n^{0.4} .\)
Table \ref{tab:mc-design-events} summarises the generator cells used in the simulation design and the corresponding realised mean event percentages at each sample size. The LLPH\(\to\)BPPH degree grid, also at \(n=500\), is in ~\ref{app:mc-tables}.

We evaluate relative bias (\%) and coverage probability
of nominal 95\% intervals, together with the \emph{standard error (SE)
calibration ratio}. Relative bias is
{\begin{align}\label{eqn:relbias}\text{relative bias (\%)}=100\times\frac{\hat\beta-\beta_0}{\beta_0},\end{align}\normalsize}
where \(\beta_0\) is the data-generating coefficient. Because $\beta_{01}$ is negative, a positive relative bias for the continuous
covariate corresponds to overestimation of the coefficient in absolute value
(a more negative estimate);
for the binary covariate, whose data-generating value is positive, the sign has
its usual interpretation. The SE calibration ratio is

{\footnotesize\begin{align}\text{SE calibration ratio}=\frac{\overline{\text{reported SE}}}{\text{Monte Carlo standard deviation}},\end{align}\normalsize}
where \(\overline{\text{reported SE}}\) is the mean of the reported
standard errors across replicates and the Monte Carlo standard deviation
is the empirical standard deviation of the coefficient estimates. Ratios
near \(1\) indicate well-calibrated uncertainty, values below \(1\) indicate
underestimation, and values above \(1\) indicate overestimation. The
summaries are computed from the data-link combinations in Table
\ref{tab:mc-design-events}.

\subsection{MLE--Bayes agreement and finite-sample calibration with a small polynomial degree}\label{sec:sim-results}

Table~\ref{tab:bp-mcsim-abc} has two purposes. First, it compares
maximum likelihood and Bayesian estimation under the weakly informative
priors of Section \ref{sec:priors}. Second, it assesses finite-sample
relative bias (\%), coverage, and SE calibration across the generator-model combinations.
Close agreement between the MLE and Bayes columns is consistent with
Bayesian fits behaving similarly to MLE while retaining mild
regularisation against implausible coefficient or baseline-parameter
values. Markov chain Monte Carlo (MCMC) settings for the Bayesian replicates use the
\texttt{spbp} defaults (4 chains per fit, 2000 iterations per chain with
1000 post-warmup draws) documented in Section~\ref{sec:mcmc}.

At \(n=100\), \(n=200\), and \(n=500\), the Weibull cells and LLPO are close to nominal coverage,
MLE and Bayes agree, and relative bias is small. This is the intended
package demonstration: when the fitted PH, PO, or AFT model matches
the data-generating regression structure, the unified
interface recovers regression coefficients with usable finite-sample
behaviour. Close MLE--Bayes agreement is consistent with limited prior
influence on the standardised scale; the weakly informative priors still
regularise implausibly large coefficient or baseline-parameter values
without materially shifting the point estimates.

Bayes coverage is modestly higher in some PO and AFT cells at \(n=50\),
where SE ratios are closer to 1, but the prior does not close large
calibration gaps. When both methods undercover, as in
LLPH\(\to\)BPPH for the continuous covariate (75\% MLE and 78\% Bayes at
\(n=100\); 89\% and 88\% at \(n=500\) with the same \(m=n^{0.4}\)),
the SE ratio below 1 points to underestimated uncertainty
rather than large relative bias. That cell is a unimodal-hazard stress
case for an uncentred sieve \citep{Osman:2012,de2025degree}. Transformed
Bernstein priors centred on a parametric family
\citep{Chen:2014,Zhou:2018} avoid this by construction; \pkg{spsurv} does
not, by design.

\begin{table*}[!h]
\centering
\scriptsize
\setlength{\tabcolsep}{1.8pt}
\renewcommand{\arraystretch}{1.05}
\caption{\label{tab:bp-mcsim-abc}
MLE--Bayes agreement and finite-sample performance (Bayes \(R=1000\); MLE complete cases; \(m=n^{0.4}\)). Coverage: 95\% intervals (Wald MLE, HPD Bayes). Relative bias (\%): \(100(\hat\beta-\beta_0)/\beta_0\). SE ratio: mean reported SE divided by Monte Carlo SD. Generator labels: W = Weibull, LL = log-logistic; PH, PO, AFT. Columns \(50\), \(100\), \(200\), and \(500\) are sample sizes \(n\).}
\begin{tabular}{@{}lll*{24}{r}@{}}
\toprule
&&&
\multicolumn{8}{c}{Coverage (\%)} &
\multicolumn{8}{c}{Rel.\ bias (\%)} &
\multicolumn{8}{c}{SE ratio} \\
\cmidrule(lr){4-11}
\cmidrule(lr){12-19}
\cmidrule(l){20-27}
&&&
\multicolumn{4}{c}{MLE} & \multicolumn{4}{c}{Bayes} &
\multicolumn{4}{c}{MLE} & \multicolumn{4}{c}{Bayes} &
\multicolumn{4}{c}{MLE} & \multicolumn{4}{c}{Bayes} \\
\cmidrule(lr){4-7}
\cmidrule(lr){8-11}
\cmidrule(lr){12-15}
\cmidrule(lr){16-19}
\cmidrule(lr){20-23}
\cmidrule(l){24-27}
Generator & Fit & & 50 & 100 & 200 & 500 & 50 & 100 & 200 & 500 & 50 & 100 & 200 & 500 & 50 & 100 & 200 & 500 & 50 & 100 & 200 & 500 & 50 & 100 & 200 & 500 \\
\midrule
WPH  & BPPH & Cont. (age) & 84.9 & 87.9 & 90.4 & 92.8 & 94.3 & 91.6 & 91.7 & 92.8 &  9.6 &  5.7 &  3.8 &  1.8 &  3.4 &  3.2 &  2.9 &  1.6 & 0.86 & 0.88 & 0.92 & 0.95 & 1.02 & 0.93 & 0.94 & 0.95 \\
WPH  & BPPH & Bin. (sex) & 91.4 & 92.6 & 93.6 & 94.0 & 94.0 & 94.4 & 93.7 & 93.6 & 10.0 &  7.1 &  4.0 &  1.5 &  4.0 &  4.7 &  3.1 &  1.3 & 0.88 & 0.93 & 0.98 & 0.99 & 0.96 & 0.96 & 0.99 & 0.99 \\
WPO  & BPPO & Cont. (age) & 93.2 & 93.5 & 94.4 & 94.6 & 95.0 & 94.5 & 96.1 & 95.7 &  3.4 &  2.5 &  2.2 &  1.1 & -4.3 & -1.8 & -0.1 &  0.1 & 0.95 & 0.95 & 0.96 & 0.98 & 1.09 & 1.03 & 1.01 & 1.01 \\
WPO  & BPPO & Bin. (sex) & 94.2 & 95.5 & 94.4 & 94.4 & 95.8 & 95.5 & 95.4 & 95.1 &  4.9 &  4.6 &  2.5 &  0.6 & -3.0 &  0.5 &  0.0 & -0.4 & 0.95 & 0.96 & 1.00 & 0.99 & 1.02 & 0.99 & 1.03 & 1.00 \\
WAFT  & BPAFT & Cont. (age) & 83.1 & 91.2 & 91.6 & 93.9 & 85.2 & 90.5 & 90.9 & 94.8 & -6.2 & -2.4 & -0.9 &  0.0 & -7.3 & -3.4 & -1.6 & -0.3 & 0.85 & 0.94 & 1.01 & 0.96 & 0.97 & 0.98 & 0.95 & 0.98 \\
WAFT  & BPAFT & Bin. (sex) & 93.3 & 95.8 & 95.6 & 94.8 & 96.5 & 96.3 & 95.5 & 96.2 & -7.5 & -4.1 & -0.6 &  0.2 & -8.1 & -4.8 & -1.2 & -0.6 & 0.99 & 1.06 & 1.11 & 0.30 & 1.10 & 1.09 & 1.05 & 0.48 \\
\addlinespace[2pt]
LLPH  & BPPH & Cont. (age) & 68.9 & 75.1 & 83.2 & 88.8 & 77.2 & 78.0 & 83.0 & 87.9 &  4.8 &  5.4 &  5.0 &  2.8 & -0.5 &  3.4 &  4.9 &  3.2 & 0.59 & 0.64 & 0.76 & 0.90 & 0.72 & 0.71 & 0.78 & 0.88 \\
LLPH  & BPPH & Bin. (sex) & 89.0 & 89.8 & 90.5 & 93.5 & 92.1 & 90.4 & 90.8 & 92.7 &  5.8 &  6.8 &  5.3 &  2.5 &  0.6 &  5.1 &  5.2 &  2.9 & 0.82 & 0.83 & 0.91 & 0.97 & 0.90 & 0.86 & 0.91 & 0.96 \\
LLPO  & BPPO & Cont. (age) & 87.6 & 89.0 & 91.4 & 92.9 & 95.2 & 92.6 & 93.4 & 93.1 &  7.2 &  4.4 &  3.4 &  1.7 &  0.1 &  1.8 &  2.6 &  1.7 & 0.86 & 0.87 & 0.91 & 0.97 & 1.03 & 0.94 & 0.95 & 0.97 \\
LLPO  & BPPO & Bin. (sex) & 93.1 & 94.9 & 95.1 & 94.5 & 95.6 & 95.2 & 94.9 & 94.1 & 10.1 &  6.8 &  3.2 &  1.0 &  2.0 &  3.9 &  2.5 &  1.0 & 0.90 & 0.93 & 1.01 & 0.98 & 0.97 & 0.96 & 1.02 & 0.99 \\
LLAFT  & BPAFT & Cont. (age) & 86.1 & 90.5 & 91.0 & 93.2 & 89.9 & 90.3 & 90.1 & 91.5 & -5.0 & -1.7 & -1.2 & -0.5 & -7.9 & -4.0 & -2.8 & -1.4 & 0.88 & 0.90 & 0.94 & 0.93 & 1.08 & 1.02 & 0.98 & 0.97 \\
LLAFT  & BPAFT & Bin. (sex) & 91.0 & 91.7 & 93.6 & 94.8 & 95.3 & 94.0 & 94.1 & 95.0 & -7.5 & -3.9 & -1.3 & -1.1 & -9.2 & -6.4 & -2.7 & -1.1 & 0.90 & 0.94 & 0.97 & 0.45 & 1.07 & 1.01 & 1.00 & 1.01 \\
\bottomrule
\end{tabular}\\[2pt]
\end{table*}

\subsubsection{Sensitivity to polynomial degree}\label{sensitivity-to-polynomial-degree}\label{sec:degree}

The package default is \(m=\sqrt{n}\) \citep{Osman:2012}, when \texttt{degree} is omitted in
\texttt{spbp.default()}. This is a
practical starting value rather than a data-adaptive optimum.
~\ref{app:mc-tables} shows that, in the LLPH\(\to\)BPPH cell,
raising \(m\) from \(n^{0.2}\) toward \(\sqrt{n}\) moves Wald coverage
and the SE ratio towards nominal, while further bias reduction is modest.
At \(n=500\), \(m=n^{0.2}\) collapses (age coverage 7\%) and
\(m=n^{0.8}\) turns down again (88\%). Users should treat \(m=\sqrt{n}\) as a
starting value rather than ``larger is safer,'' and increase it if interval
calibration looks poor. A systematic study of shape versus degree is beyond
the scope of this software paper.

\section{Motivating examples}\label{sec:illustrations}

\subsection{Laryngeal cancer}\label{sec:illus-larynx}

\citet{Kardaun:1983} reports a cohort of patients with laryngeal cancer treated
in Dutch hospitals in the 1970s. We examine whether survival is
associated with age at diagnosis and tumour stage (I--IV). The \texttt{larynx} data in \pkg{KMsurv} \citep{Klein:1997} contain
\texttt{time} (years to death or censoring), \texttt{delta} (event
indicator), \texttt{age} (years), \texttt{stage} (1--4), and \texttt{diagyr}
(year of diagnosis). The data set has
\(n=90\) subjects with 50 deaths. With default treatment contrasts,
stage effects compare each level to Stage I; the continuous coefficient is
the log hazard ratio per additional year of age, holding stage fixed.
Proportional hazards models are a natural starting point for staged
oncology outcomes; we therefore fit a BPPH model by both maximum likelihood
and Bayesian approaches, comparing results with Cox partial likelihood
estimates.

\subsubsection{Model fit and coefficients}\label{model-fit-and-coefficients}

The BPPH model described in Section~\ref{sec:models} was fitted to the
\texttt{larynx} data using maximum likelihood estimation. Categorical
covariates such as \texttt{stage} are converted to factors in the data
before model fitting rather than through \texttt{factor()} in the formula,
so coefficient labels appear as \texttt{stage2}, \texttt{stage3}, and
\texttt{stage4}. Numeric output in the examples below is rounded to two
decimal places for display. The corresponding hazard-ratio estimates are
reported below.

{\footnotesize\begin{verbatim}
data("larynx", package = "KMsurv")
larynx$stage <- factor(larynx$stage)

bpph_fit <- bpph(
  Surv(time, delta) ~ age + stage,
  data = larynx,
  approach = "mle"
)
\end{verbatim}}

{\footnotesize\begin{verbatim}
tidy(bpph_fit)
\end{verbatim}}

{\footnotesize\begin{verbatim}
#>     term component estimate std.error statistic p.value
#> 1    age      coef     0.02      0.01      1.34    0.18
#> 2 stage2      coef     0.17      0.46      0.37    0.71
#> 3 stage3      coef     0.66      0.36      1.85    0.06
#> 4 stage4      coef     1.80      0.43      4.19  2.75e-05
\end{verbatim}}

{\footnotesize\begin{verbatim}
glance(bpph_fit)
\end{verbatim}}

{\footnotesize\begin{verbatim}
#>    n nevent  logLik approach model df statistic
#> p.value  rsq max.rsq    AIC    BIC
#> 1 90     50 -140.05      mle    ph 14     19.57
#> 6.07e-04 0.20    0.96 308.10 343.10
\end{verbatim}}

\needspace{6\baselineskip}

The estimated hazard ratio for Stage IV versus Stage I at the same age
is 6.05, with 95\% limits 2.61
and 14.02 from \texttt{tidy(bpph\_fit, conf.int = TRUE, exponentiate = TRUE)},
indicating higher mortality for Stage IV than for Stage
I. The \texttt{tidy()} and \texttt{glance()} methods from \pkg{generics} provide
broom-style summaries. For maximum-likelihood fits, \texttt{glance()} and \texttt{AIC()}
count both regression and Bernstein (\texttt{gamma}) parameters
(\texttt{.spbp\_nparams()}); the likelihood-ratio \texttt{df} in
\texttt{summary()} remains the number of regression terms only.
The \texttt{print()} and \texttt{summary()} methods provide conventional
coefficient, interval, and model-level summaries.

\subsubsection{Survival prediction}\label{survival-prediction}

Predicted survival curves for \texttt{spbp} objects are obtained with
\texttt{survfit.spbp}. Figure~\ref{fig:larynx-survfit-plot} illustrates
maximum-likelihood and Bayesian survival prediction under different
Bernstein degrees, including the effect of an ill-conditioned
Bernstein-parameter information block on delta-method uncertainty.

The method also returns median-survival estimates and associated limits
when the relevant survival-curve crossings are identified within the
observed follow-up.
Survival decreases with
increasing stage at age 65 years (rounded from the sample mean age
64.61 years). Where a curve crosses 0.5, that time is an estimated
median survival on the time scale of the data; curves that remain above 0.5 on the plotted
follow-up have no median crossing visible on the grid shown.

{\footnotesize\begin{verbatim}
nd <- data.frame(
  age   = 65,
  stage = factor(levels(larynx$stage))
)

survfit(bpph_fit, newdata = nd)
\end{verbatim}}

Figure \ref{fig:larynx-survfit-plot}~Panel~(a) uses the default Bernstein degree
$m=\sqrt{n}=10$ on the 90-patient \texttt{larynx} sample;
the $\boldsymbol{\gamma}$ information block is then ill-conditioned
(Section~\ref{sec:delta-method}), so \texttt{survfit.spbp} reports pointwise
delta-method bands together with a warning that they may be unreliable
(printed at this step only).
Panel~(b) contrasts the same covariate profiles under three fits:
$m=3$ maximum likelihood (stable $\boldsymbol{\gamma}$ information and
95\% delta-method bands without an instability warning), default
$m=10$ maximum likelihood (ill-conditioned $\boldsymbol{\gamma}$ block and
unreliable delta-method bands with a \texttt{survfit()} warning), and
default $m=10$ Bayesian (posterior HPD bands at the same degree, without
the delta method).
Lowering $m$ trades baseline flexibility for identifiability on small
samples and is one adjustment suggested by the \texttt{survfit()} warning;
panel~(b) shows that refitting with \texttt{approach = "bayes"} at the
same degree is the other main route for curve-wise uncertainty. The next
subsection gives posterior summaries and code for the Bayesian fit.

\begin{figure*}[!h]
\centering
\begin{minipage}{0.48\linewidth}
\centering
\includegraphics[width=\linewidth]{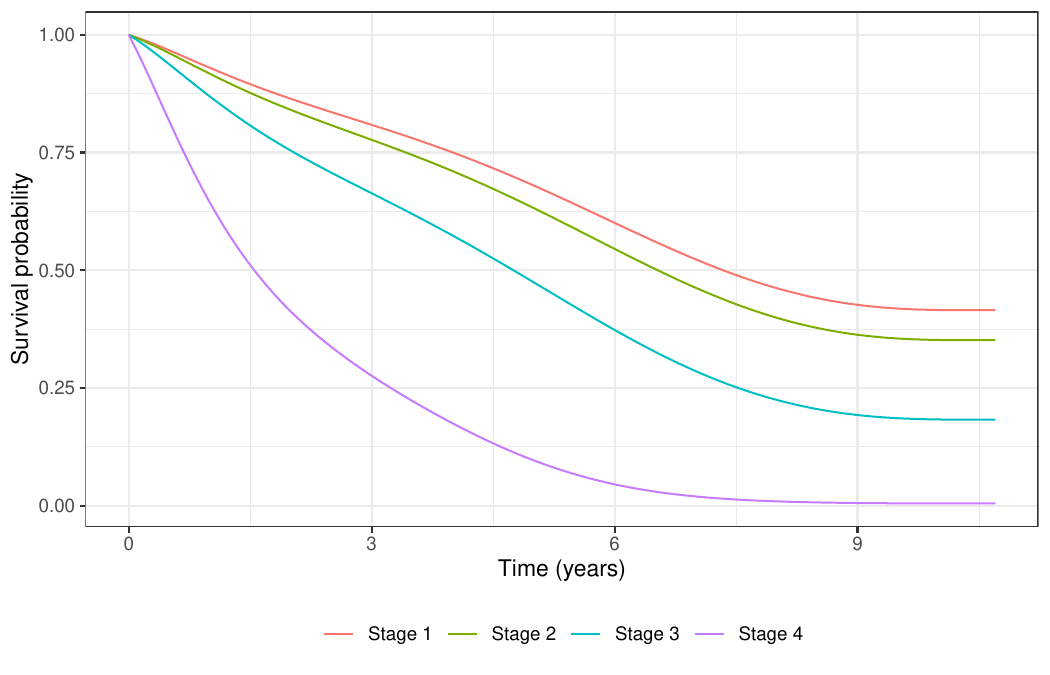}\\[0.4em]
{\footnotesize (a) Default $m=10$ MLE with delta-method bands.}
\end{minipage}\hfill
\begin{minipage}{0.48\linewidth}
\centering
\includegraphics[width=\linewidth]{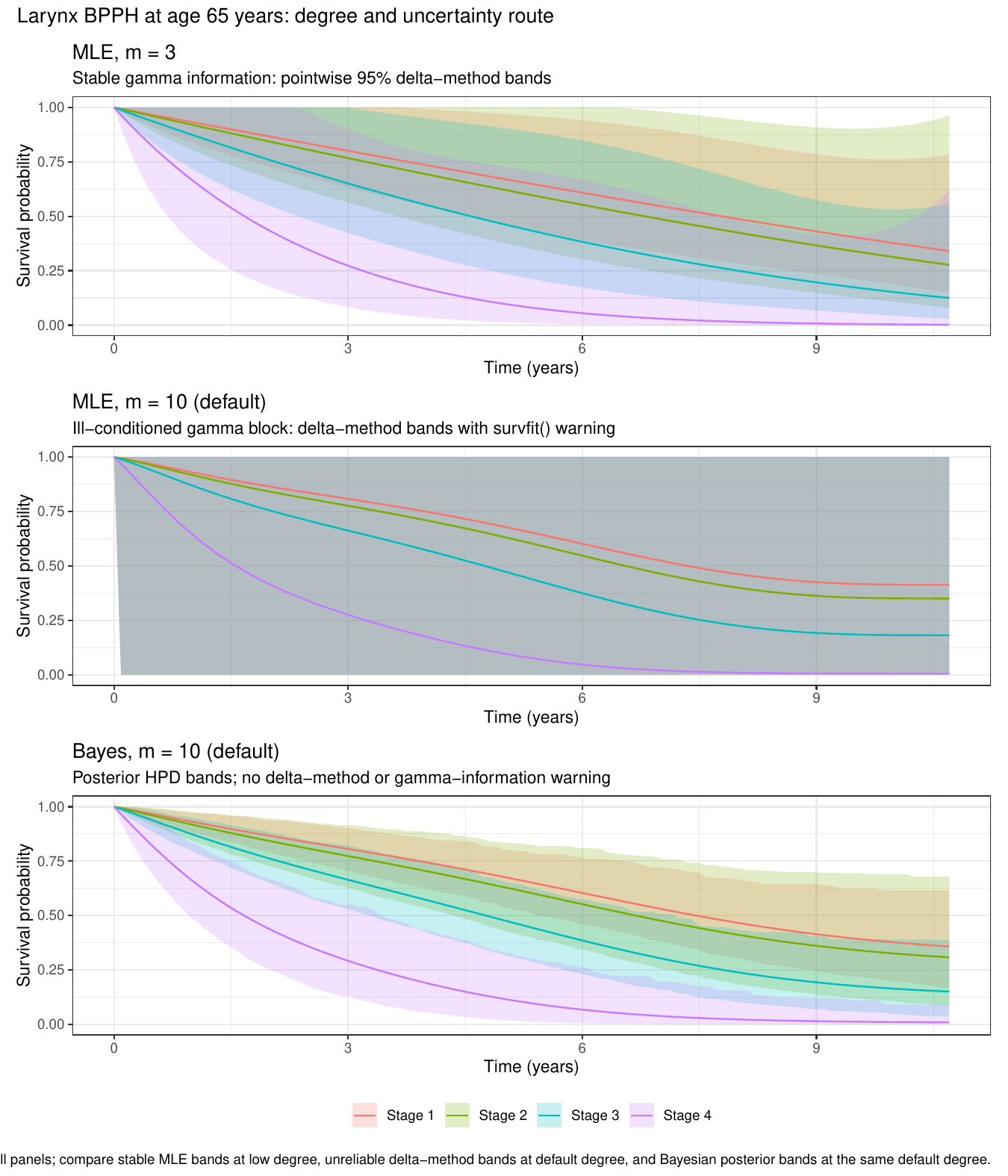}\\[0.4em]
{\footnotesize (b) $m=3$ MLE, $m=10$ MLE, and $m=10$ Bayes (top to bottom).}
\end{minipage}
\caption{Estimated BPPH survival curves for patients aged 65 years in the
larynx data. Panel~(a) shows the default \(m=10\) maximum-likelihood fit.
Panel~(b) compares \(m=3\) maximum likelihood, \(m=10\) maximum
likelihood, and \(m=10\) Bayesian fits. Shaded regions show pointwise
95\% delta-method intervals for maximum-likelihood fits and HPD
credible intervals for the Bayesian fit. Survival decreases from
Stage~I to Stage~IV.}
\label{fig:larynx-survfit-plot}
\end{figure*}

\subsubsection{Bayesian analysis}\label{bayesian-analysis}

As an alternative to lowering Bernstein \texttt{degree} when maximum-likelihood
delta-method bands are unreliable (Figure~\ref{fig:larynx-survfit-plot},
panel~(b)), the same model can be refitted with
\texttt{approach\ =\ "bayes"}; the posterior
summaries agree closely with the maximum likelihood estimates, as the
priors are weakly informative by default, using the prior specification
of Section \ref{sec:priors}. Unlike the maximum-likelihood fit above,
the Bayesian \texttt{survfit.spbp} summary returns finite credible limits
for median survival when the posterior supports them. Curve-wise bands
use posterior draws of \((\boldsymbol{\beta},\boldsymbol{\gamma})\) with
\texttt{interval.type = "hpd"} by default and optional monotone enforcement
over time. Bayesian fits use the package's default MCMC settings; the
complete computational settings are reported in
Section~\ref{sec:mcmc}.

{\footnotesize\begin{verbatim}

set.seed(1)

bpph_fit_bayes <- bpph(
  Surv(time, delta) ~ age + stage,
  data = larynx,
  approach = "bayes"
)

survfit(bpph_fit_bayes, newdata = nd)

plot_times <- seq(0, max(larynx$time), length.out = 121)

sf_bayes <- survfit(
  bpph_fit_bayes,
  newdata = nd,
  times = plot_times,
  type = "log",
  interval.type = "hpd"
)
\end{verbatim}}

{\footnotesize\begin{verbatim}
#> Call: survfit.spbp(formula = bpph_fit_bayes, newdata = nd)
#>
#>    n events median 0.95LCL 0.95UCL
#> 1 90     50   7.46    5.51      NA
#> 2 90     50   6.70    4.43      NA
#> 3 90     50   4.86    3.35    7.24
#> 4 90     50   1.73    0.97    3.57
\end{verbatim}}

\needspace{6\baselineskip}

Posterior summaries provide highest posterior density (HPD) intervals
and model comparison via deviance information criterion
\citep{spiegelhalter2002bayesian} and widely applicable information
criterion (WAIC) \citep{Vehtari:2017, watanabe2013widely} when available,
with the same PH interpretation. These quantities are reported by
\texttt{tidy()} and \texttt{glance()}; \texttt{glance()} also collects
log pseudo-marginal likelihood
\citep{geisser1979predictive, ibrahim2001bayesian} when the posterior
log-likelihood is available.

{\footnotesize\begin{verbatim}
tidy(bpph_fit_bayes)
\end{verbatim}}

{\footnotesize\begin{verbatim}
#>     term component estimate std.error
#> 1    age      coef     0.02      0.01
#> 2 stage2      coef     0.15      0.47
#> 3 stage3      coef     0.65      0.35
#> 4 stage4      coef     1.79      0.41
\end{verbatim}}

{\footnotesize\begin{verbatim}
glance(bpph_fit_bayes)
\end{verbatim}}

{\footnotesize\begin{verbatim}
#>    n nevent  logLik approach model df    waic    dic    lpml
#> 1 90     50 -144.60    bayes    ph  4 -149.21 296.95 -149.33
\end{verbatim}}

\texttt{coef()},
\texttt{model.matrix()}, and \texttt{credint()} follow
the conventions of standard fitted objects and integrate with the
\pkg{generics} / broom workflow. When comparing models,
posterior means and HPD limits from \texttt{tidy()} can be read alongside
WAIC and log pseudo-marginal likelihood from \texttt{glance()}, for example when
choosing between PO and AFT formulations that both fit the data reasonably well.

\subsubsection{Residual diagnostics}\label{residual-diagnostics}

Having compared the maximum-likelihood and Bayesian BPPH fits directly,
we now check both against an external benchmark: a Cox model with the
same linear predictor. Martingale residuals depend on the estimated
survival function, so we extract \texttt{residuals()} from each fit with
matching formulae. Martingale residuals from the maximum-likelihood and Bayesian BPPH fits
were closely aligned with those from the corresponding Cox model under
the same linear predictor, supporting similar fitted hazard structure.
Overall, the BPPH specification is a flexible alternative to
partial likelihood when PH is plausible.

{\footnotesize\begin{verbatim}
mod_cox <- coxph(
  Surv(time, delta) ~ age + stage,
  data = larynx
)

resid_df <- data.frame(
  ml_bpph = residuals(bpph_fit, type = "martingale"),
  bayes_bpph = residuals(bpph_fit_bayes,
    type = "martingale"),
  cox = residuals(mod_cox, type = "martingale")
)
\end{verbatim}}

\subsection{Veteran lung cancer (no prior therapy)}\label{sec:illus-veteran}

The Veterans Administration lung cancer trial \citep{Prentice:1973,kalbfleisch2002statistical} enrolled
patients with advanced, inoperable lung cancer. Following \citet{Pettitt:1984}, we restrict
analysis to participants with \texttt{prior\ ==\ 0} (no prior therapy), as in
that proportional-odds reanalysis of the trial. The
\texttt{veteran} data in \pkg{survival} \citep{Therneau:2020} provide survival
\texttt{time} (days) and censoring \texttt{status}, Karnofsky performance score
\texttt{karno} (0--100), histologic celltype \texttt{celltype}, treatment \texttt{trt}, and other
covariates. Following \citet{Bennett:1983}, we model \texttt{karno} and \texttt{celltype}
only (omitting \texttt{trt}, \texttt{age}, and \texttt{diagtime}); that paper also questions
the proportional hazards assumption for this subset, so we fit Bayesian
BPPO and BPAFT as semi-parametric alternatives to proportional hazards.
The restricted subsample has \(n=97\) subjects with 91 deaths.

\subsubsection{Alternatives to proportional hazards}\label{sec:alternatives}

We fit Bayesian BPPO and BPAFT models to the \texttt{veteran2} subsample
with \texttt{bppo()} and \texttt{bpaft()}, whose likelihood structures are given in
Table \ref{tab:models}, and summarise the regression parameters.
Histology uses large cell as the reference level; \texttt{karno} enters
linearly, so its coefficient is the change in the linear predictor per
one-point increase on the Karnofsky scale, holding cell type fixed.
Predicted survival curves below use Karnofsky scores 30 and 70 with
squamous histology, matching the \texttt{newdata2} object passed to
the method \texttt{survfit.spbp}. These two scores were chosen as contrasting clinical
profiles (low versus higher functional status) to make survival
differences easier to compare.

{\footnotesize\begin{verbatim}
data("veteran", package = "survival")

# Pettitt (1984) analytic subsample: no prior therapy
veteran2 <- veteran |>
  dplyr::filter(prior == 0) |>
  dplyr::mutate(celltype = factor(
    celltype,
    levels = c("large", "adeno", "smallcell", "squamous")
  ))

f <- Surv(time, status) ~ karno + celltype

fit_po  <- bppo(f, data = veteran2, approach = "bayes")
fit_aft <- bpaft(f, data = veteran2, approach = "bayes")
\end{verbatim}}

BPAFT can be more computationally demanding than BPPH and BPPO because
the residual domain depends on the regression coefficients and must be
updated during estimation (Section~\ref{sec:bp-domain}). Bayesian BPAFT
fits can benefit from standard No-U-Turn Sampler (NUTS) diagnostic checks; users should treat
divergent transitions and low effective sample sizes as diagnostics
suggesting further tuning rather than negligible numerical variation.

Posterior mean survival for squamous histology at Karnofsky scores
30 and 70 is obtained with \texttt{survfit.spbp} for each
Bayesian fit.

\subsubsection{Survival comparison}\label{survival-comparison}

Figure \ref{fig:bpaftxbppo} is generated by calling \texttt{survfit.spbp}
(in \pkg{spsurv}) on the Bayesian BPPO and BPAFT fits using \texttt{newdata2},
which fixes histology at squamous cell type and sets two contrasting clinical
profiles (30 and 70). The resulting objects are tidied with
\pkg{ggsurvfit} \citep{ggsurvfit:2024} (\texttt{tidy\_survfit()}) and plotted
in \pkg{ggplot2} with pointwise HPD credible bands from posterior simulation.

\begin{figure}[!h]
\includegraphics[width=1\linewidth]{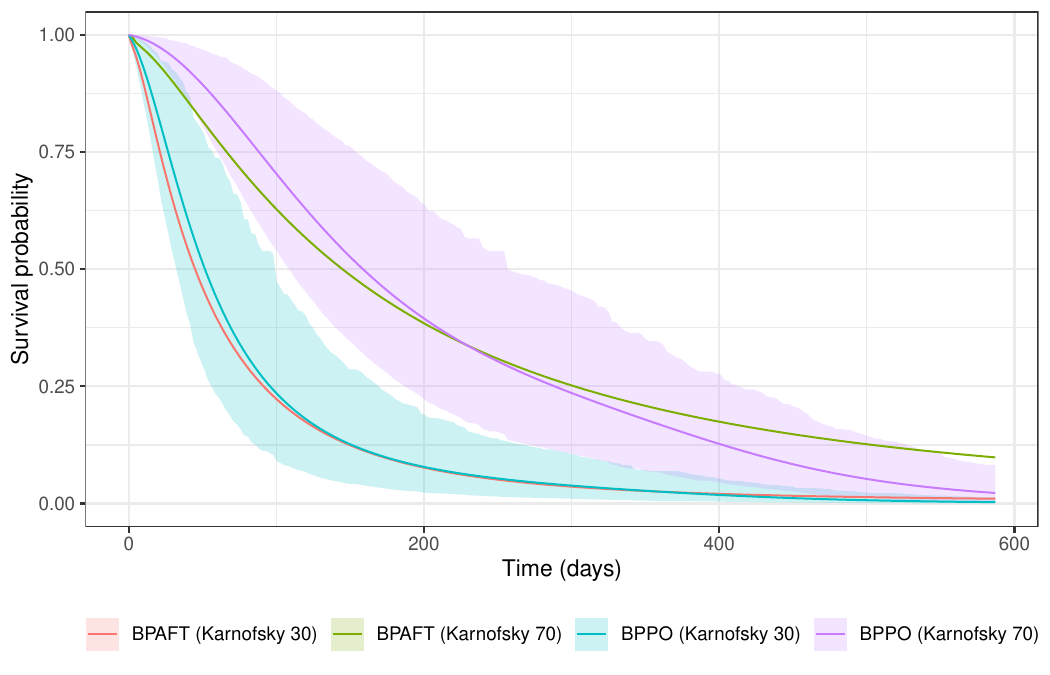} \caption{Veteran no-prior-therapy subset: Bayesian BPPO and BPAFT posterior mean survival for squamous histology at Karnofsky 30 and 70 (\texttt{newdata2}). Pointwise 95\% HPD bands from \texttt{survfit.spbp}; within each model, lower Karnofsky (30) has lower survival.}\label{fig:bpaftxbppo}
\end{figure}

Figure \ref{fig:bpaftxbppo} compares posterior mean survival for
squamous histology at Karnofsky scores 30 and 70.
Within each performance-status stratum, the BPPO and BPAFT mean curves
are similar and the pointwise credible bands overlap substantially.
Accordingly, when proportional hazards is doubtful, BPPO and BPAFT provide semi-parametric alternatives that relax proportional hazards. The overlap of their survival curves should not be read as a general property of the BP formulation. The log-logistic distribution is the only family belonging to both the proportional-odds and the accelerated-failure-time classes, and it provides a satisfactory fit to these data; the agreement between BPPO and BPAFT here therefore reflects the fact that the Bernstein basis is recovering a baseline close to a log-logistic one, under which the two regression structures coincide. That is also why LLPH, not LLPO or LLAFT, was the hard cell in Section~\ref{sec:simulation}: a log-logistic hazard with shape \(>1\) is unimodal, and PH is not a native log-logistic family. In data sets whose baseline is not well approximated by a log-logistic form, PO and AFT fits need not agree, and the choice between them should be guided by the diagnostics and model-comparison summaries described below rather than assumed.




\subsubsection{Model fit diagnostics}\label{model-fit-diagnostics}

Diagnostics follow standard survival-analysis practice \citep{survival:2000, Collett2015}. For \pkg{spsurv} objects, \texttt{residuals()} accepts
\texttt{type\ =\ "martingale"} (default), \texttt{"deviance"}, and \texttt{"cox-snell"}.
Under a well-specified model, a Kaplan--Meier estimate of the
cumulative hazard of Cox--Snell residuals is expected to lie close to
the identity line. We compute Cox--Snell residuals with
\texttt{residuals(...,\ type\ =\ "cox-snell")} and plot the cumulative hazard of
those residuals against the reference.
Deviance residuals transform martingale residuals to a signed square-root
scale that emphasises subjects poorly explained by the fitted model; they are
available via \texttt{residuals(...,\ type\ =\ "deviance")} and are useful for
flagging influential or outlying observations after global calibration has been
checked.

\begin{figure}[!h]
\includegraphics[width=1\linewidth]{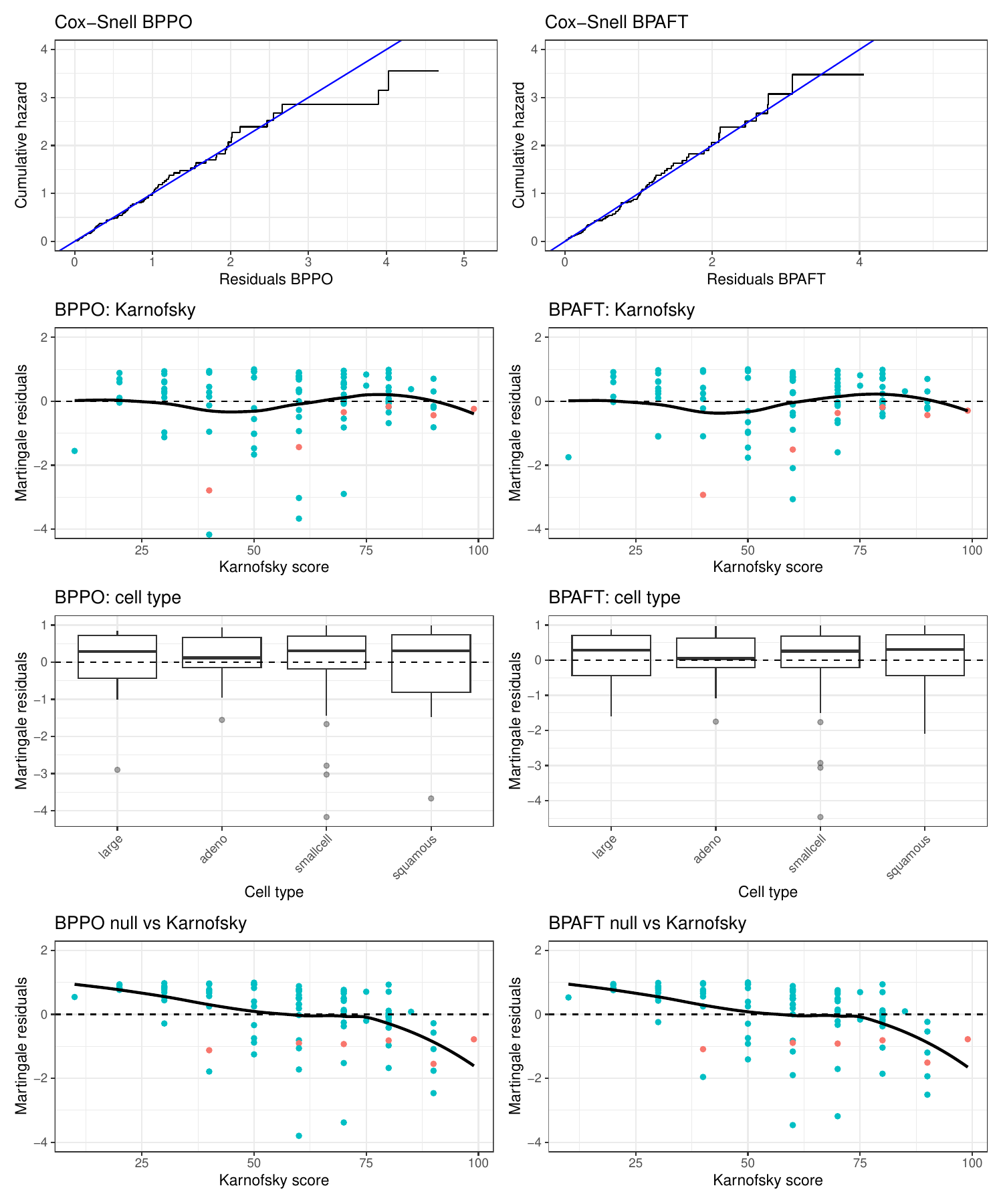} \caption{BPPO and BPAFT diagnostics (veteran no-prior-therapy). Top: Cox--Snell cumulative hazard vs 45° reference (global fit). Rows~2--3: full-model martingale residuals vs Karnofsky and cell type (functional form).}\label{fig:martingale}
\end{figure}

For covariate functional form, martingale residuals should be taken from
the fitted model that includes all covariates in the linear predictor.
Scatter plots with LOESS smooths of full-model martingale residuals
against each continuous covariate assess possible nonlinearity after
adjustment; boxplots against categorical covariates summarise residual
patterns within factor levels. Under a well-specified linear effect, the
smooth against a continuous covariate should be approximately flat around
zero.

{\footnotesize\begin{verbatim}
mod_cox_vet <- coxph(
  Surv(time, status) ~ karno + celltype,
  data = veteran2
)

fit_po_mle  <- bppo(f, data = veteran2, approach = "mle")
fit_aft_mle <- bpaft(f, data = veteran2, approach = "mle")

mr_bppo <- residuals(fit_po_mle, type = "martingale")
mr_bpaft <- residuals(fit_aft_mle, type = "martingale")
mr_cox <- residuals(mod_cox_vet, type = "martingale")
\end{verbatim}}

Figure~\ref{fig:martingale} (rows~2--3) shows these adjusted
diagnostics for BPPO and BPAFT. Full-model martingale patterns are similar for BPPO and BPAFT. The
adjusted smooth against Karnofsky score shows mild departure from
linearity, so a nonlinear performance status term may be worth exploring
in a full analysis; cell-type boxplots show no strong imbalance across
histology levels after fitting the full linear predictor.

\section{Discussion}\label{sec:discussion}

Semi-parametric survival modelling based on the BP was previously
introduced by \citet{Osman:2012}. The methods in the package do not require a
parametric family for the baseline; instead, they rely on the polynomial
coefficient structure. In contrast to partial-likelihood Cox regression,
where the baseline hazard is left unspecified and absorbed by the
partial likelihood, the BP formulation expresses baseline hazard, odds,
or log-time features explicitly through polynomial coefficients and
estimates smooth baseline shapes via this sieve construction, while
retaining the usual PH/PO/AFT regression interpretation for covariate
effects. AFT uses the monomial construction on the adaptive log-time
domain, at the added per-iteration cost discussed in
Section~\ref{sec:bp-domain}.

When there is no strong scientific reason to prefer one regression
family, we recommend fitting at least BPPO and BPAFT, with BPPH added
when a proportional-hazards interpretation is important. Model choice
should then be informed by predicted survival, residual diagnostics, and
model-comparison summaries \citep[Chapter 7]{kalbfleisch2002statistical}.
Similar conclusions across families provide evidence that the
substantive result is not highly sensitive to model choice; substantial
differences should be reported and investigated.

In the aligned Weibull cells that represent routine use, Bernstein PH,
PO, and AFT fits recover regression effects with small relative bias and
near-nominal coverage at \(n=100\), \(n=200\), and \(n=500\), and maximum-likelihood and Bayesian
summaries agree. The default weakly informative priors mainly regularise.
Interval calibration is poorer for the log-logistic PH generator, where
a unimodal baseline is hard for a small uncentred degree; raising \(m\)
toward \(\sqrt{n}\) improves coverage, while a much larger degree can turn
down again at \(n=500\). We therefore recommend the package defaults
(\(m=\sqrt{n}\), internal standardisation, weakly informative
gamma/normal priors) as a first fit, with a local degree check when
calibration is a concern.

\section{Computational details}\label{sec:computational}

The results in this paper were obtained using R 4.5.1 and \pkg{spsurv} 1.1.0
\citep{spsurv:2026}, with maximum-likelihood fits relying on \pkg{MASS} 7.3-65
and Bayesian fits obtained with \pkg{rstan} 2.32.7 (Stan 2.32.2;
\pkg{StanHeaders} 2.32.10). All packages used in this manuscript are available
from CRAN.

\subsection{Reproducibility}\label{sec:reproducibility}

The larynx and veteran workflows load the data sets  \texttt{KMsurv::larynx}
and \texttt{survival::veteran}, factor \texttt{stage} or \texttt{celltype},
and fit \pkg{spsurv} models with default \texttt{degree} (when omitted,
\(m=\sqrt{n}\)), \texttt{scale = TRUE}, and \texttt{chains = 4}
for Bayesian examples (\texttt{spbp.default()} default). Predicted survival from
Bernstein fits is computed with the \texttt{survfit.spbp}; multi-curve objects
are tidied with \pkg{ggsurvfit} \citep{ggsurvfit:2024} and plotted with
\pkg{ggplot2}. The Bayesian larynx fit uses
\texttt{set.seed(1)} before calling \texttt{bpph(..., approach = "bayes")}.
Monte Carlo performance measures are defined in
Section~\ref{sec:simulation}.

\subsubsection{Bayesian computation and MCMC settings}\label{sec:mcmc}

Bayesian estimation uses \texttt{rstan::sampling()}. The clinical
illustrations (Section~\ref{sec:illustrations}) and the Monte Carlo
simulation study (Section~\ref{sec:simulation}) use the settings
reported in Table~\ref{tab:mcmc-settings}. Users may increase
\texttt{adapt\_delta} when NUTS diagnostics indicate problems.
The degree-sensitivity study (Figure~\ref{fig:degree-llph-bpph-plot}) used
MLE only.

\begin{table}[!h]
\footnotesize
\centering
\setlength{\tabcolsep}{4pt}
\renewcommand{\arraystretch}{1.1}
\caption{\label{tab:mcmc-settings}
Bayesian MCMC settings used for the clinical illustrations  and the Monte Carlo simulation study.}
\begin{tabularx}{\columnwidth}{@{}
  >{\raggedright\arraybackslash}p{0.32\columnwidth}
  >{\raggedright\arraybackslash}X
  >{\raggedright\arraybackslash}X
@{}}
\toprule
 & Clinical illustrations & Monte Carlo simulation \\
\midrule
Chains & 4 & 4 \\
Iterations per chain & 2000 & 2000 \\
Warm-up iterations & 1000 & 1000 \\
Post-warm-up draws retained & 1000 (no thinning) & 1000 (no thinning) \\
Random seed & \texttt{set.seed(1)} (larynx fit) & \texttt{set.seed(r)}, \(r=1,\ldots,R\), \(R=1000\) \\
Other settings & \texttt{cores} from \texttt{spbp.default()} & \texttt{cores} from \texttt{spbp.default()} \\
\bottomrule
\end{tabularx}
\end{table}

\section{Acknowledgements}\label{acknowledgements}

The authors are grateful to everyone who contributed to discussions of this R package through helpful conversations, especially Dr.\ Rumenick Pereira da Silva, Dr.\ Silvio Cabral Patrício, Prof.\ Dr.\ Dani Gamerman, Prof.\ Dr.\ Marcos Oliveira Prates, and Prof.\ Dr.\ Marcelo Azevedo Costa. The authors acknowledge scholarship support from CEMIG and support from CNPq, CAPES, and FAPEMIG in Brazil. Additional institutional support came from the Department of Statistics and the School of Engineering at the Federal University of Minas Gerais (UFMG).

\clearpage
\appendix

\section{Beta-density monomial expansion for BPAFT}\label{app:monomial}

Section~\ref{sec:bp-domain} evaluates the BPAFT baseline at the scaled
residual \(u\in[0,1]\) using the same \(m\) beta densities as BPPH and BPPO,
\(f_{\beta}(u\mid k,\,m-k+1)\) for \(k=1,\ldots,m\), rather than the
\(m+1\) Bernstein polynomials \(b_{i,m}\) indexed from \(0\) to \(m\).
Expanding \((1-u)^{m-k}\) gives the monomial form implemented by
\texttt{pw.basis()}:
{\footnotesize\begin{align}
f_{\beta}(u\mid k,\,m-k+1)
&=
\frac{u^{k-1}(1-u)^{m-k}}{\mathrm{B}(k,\,m-k+1)}
=
\sum_{j=1}^{m}
P_{jk}\,u^{j-1},
\end{align}\normalsize}
where \(\mathrm{B}(\cdot,\cdot)\) is the beta function and
{\footnotesize\begin{align}
P_{jk}
=
\frac{(-1)^{j-k}}{\mathrm{B}(k,\,m-k+1)}
\binom{m-k}{j-k},
\qquad j,k=1,\ldots,m,
\end{align}\normalsize}
with the convention that the binomial coefficient is zero when \(j<k\).
The matrix \(P=(P_{jk})\) depends only on \(m\) and is formed once per fit.
If \(\boldsymbol{p}(u)=(u^{0},\ldots,u^{m-1})\) and
\(\boldsymbol{q}(u)=(u^{1}/1,\ldots,u^{m}/m)\), the hazard and
cumulative bases on the residual scale with range
\(R=w_{\max}(\boldsymbol{\beta})-w_{\min}(\boldsymbol{\beta})\) are
\(\boldsymbol{g}(u)=\boldsymbol{p}(u)P/R\) and
\(\boldsymbol{G}(u)=\boldsymbol{q}(u)P\).
Only the powers of \(u_i(\boldsymbol{\beta})\) are recomputed when
\(\boldsymbol{\beta}\) changes during estimation (Section~\ref{sec:bp-domain}).

\section{AFT delta-method gradients and parameterisation comparison}\label{app:aft-gradients}

Section~\ref{sec:bp-domain} notes that gradients of baseline quantities with
respect to \(\boldsymbol{\beta}\) under the BPAFT rescaling include endpoint
correction terms beyond the ordinary \(-\boldsymbol{x}_i\) contribution.
This appendix gives the full expressions used by \texttt{survfit.spbp} and
compares the resulting derivative structure with an alternative
covariate-scaled-time parameterisation.

For a differentiable baseline quantity \(q\) evaluated at the scaled
argument \(u_i(\beta)\), the chain rule gives
{\footnotesize\begin{align}
\frac{\partial u_i}{\partial\boldsymbol{\beta}}
&=
\frac{1}{R(\boldsymbol{\beta})}
\left(
-\boldsymbol{x}_i
+\boldsymbol{x}_{i^\ast}
-u_i(\boldsymbol{\beta})\bigl[\boldsymbol{x}_{i^\ast}-\boldsymbol{x}_{j^\ast}\bigr]
\right),
\label{eq:aft-u-deriv}
\end{align}\normalsize}
where \(R(\boldsymbol{\beta})=w_{\max}(\boldsymbol{\beta})-w_{\min}(\boldsymbol{\beta})\),
\(i^\ast=\arg\min_i w_i(\boldsymbol{\beta})\), and
\(j^\ast=\arg\max_i w_i(\boldsymbol{\beta})\).
Only the two training subjects at the residual extremes contribute
through \(\boldsymbol{x}_{i^\ast}\) and \(\boldsymbol{x}_{j^\ast}\); all other
subjects enter only through \(-\boldsymbol{x}_i/R\).
For a predicted cumulative hazard
\(H(t\mid\boldsymbol{x})=\boldsymbol{\gamma}'\boldsymbol{G}\bigl(u(\boldsymbol{\beta})\bigr)\)
at fixed \(t\), with \(h=\partial H/\partial w\) the log-time hazard,
{\footnotesize\begin{align}
\frac{\partial H(t\mid\boldsymbol{x})}{\partial\boldsymbol{\beta}}
=
h\,
\left(
-\boldsymbol{x}
+\boldsymbol{x}_{i^\ast}
-u(\boldsymbol{\beta})\bigl[\boldsymbol{x}_{i^\ast}-\boldsymbol{x}_{j^\ast}\bigr]
\right).
\label{eq:aft-H-grad}
\end{align}\normalsize}
Equations~\eqref{eq:aft-u-deriv}--\eqref{eq:aft-H-grad} are the gradients
used in \texttt{survfit.spbp} delta-method bands for BPAFT; they reduce to
the frozen-endpoint form \(-h\,\boldsymbol{x}\) only if
\(w_{\min}\) and \(w_{\max}\) are treated as fixed at the MLE. For any differentiable baseline quantity
\(q(w_i)\) evaluated directly at the log residual (not composed with
\(u_i(\boldsymbol{\beta})\)),
{\footnotesize\begin{align}
\frac{\partial w_i}{\partial\boldsymbol{\beta}}
=
-\boldsymbol{x}_i,
\qquad
\frac{\partial q(w_i)}{\partial\boldsymbol{\beta}}
=
-q'(w_i)\,\boldsymbol{x}_i,
\qquad
\frac{\partial^2 q(w_i)}{\partial\boldsymbol{\beta}\partial\boldsymbol{\beta}'}
=
q''(w_i)\,\boldsymbol{x}_i\boldsymbol{x}_i'.
\label{eq:aft-deriv}
\end{align}\normalsize}
Thus, score contributions for \(\boldsymbol{\beta}\) can be obtained by
differentiating the BP representation once with respect to its scalar
argument and then multiplying by the covariates, and the second
derivative in \eqref{eq:aft-deriv} has the outer-product form
\(\boldsymbol{x}_i\boldsymbol{x}_i'\). This structure is simpler than
differentiating a full covariate-dependent time-scale representation.
Table~\ref{tab:aft-deriv-compare} compares the two parametrisations
directly: writing the scaled time as
\(v_i=y_i\,e^{-\boldsymbol{\beta}'\boldsymbol{x}_i}\),
so that \(w_i=\log v_i\), the same baseline quantity
can equivalently be written as \(q(w_i)\) or as
\(p\{v_i\}=q(w_i)\) on the scaled-time index. Because
the index derivative \(\partial w_i/\partial\boldsymbol{\beta}\) in
\eqref{eq:aft-deriv} is the constant \(-\boldsymbol{x}_i\), every order of
derivative on the \(w_i\) scale stops at \(\boldsymbol{x}_i\) or
\(\boldsymbol{x}_i\boldsymbol{x}_i'\); on the \(v_i\)
scale, the index derivative is itself proportional to
\(v_i\), so it is reintroduced at every derivative
order, which is the additional multiplicative time-scale structure in
the Hessian that the log-residual formulation avoids.

\begin{table}[!h]
\footnotesize
\centering
\caption{\label{tab:aft-deriv-compare} Derivative structure under log-time residual \(w_i\) vs covariate-scaled time \(v_i\) (\(w_i=\log v_i\)) for a generic baseline quantity. \(q(\cdot)\) and \(p(\cdot)\) denote baseline functions; primes denote derivatives w.r.t.\ the scalar argument.}
\begin{tabularx}{\columnwidth}{%
>{\raggedright\arraybackslash}p{0.27\columnwidth}
>{\raggedright\arraybackslash}X
>{\raggedright\arraybackslash}X}
\toprule
 & Log-time residual \(w_i\) & Scaled time \(v_i\) \\
\midrule
Index &
\(w_i=\log(y_i)-\boldsymbol{\beta}'\boldsymbol{x}_i\) &
\(v_i=y_i\,e^{-\boldsymbol{\beta}'\boldsymbol{x}_i}\)
\\[0.7em]
Baseline quantity &
\(q(w_i)\) &
\(p\{v_i\}=q(w_i)\)
\\[0.7em]
\(\partial(\text{index})/\partial\boldsymbol{\beta}\) &
\(-\boldsymbol{x}_i\) (constant) &
\(-\boldsymbol{x}_i\,v_i\) (\(\boldsymbol{\beta}\)-dependent)
\\[0.7em]
\(\partial(\text{baseline})/\partial\boldsymbol{\beta}\) &
\(-q'(w_i)\,\boldsymbol{x}_i\) &
\(-p'\{v_i\}\,v_i\,\boldsymbol{x}_i\)
\\[0.7em]
\(\partial^2(\text{baseline})/\partial\boldsymbol{\beta}\partial\boldsymbol{\beta}'\) &
\(q''(w_i)\,\boldsymbol{x}_i\boldsymbol{x}_i'\) &
\(\bigl[v_i\,p'\{v_i\}+v_i^2\,p''\{v_i\}\bigr]\boldsymbol{x}_i\boldsymbol{x}_i'\)
\\[0.7em]
Extra \(\boldsymbol{\beta}\)-dependence beyond
\(\boldsymbol{x}_i\boldsymbol{x}_i'\) &
none for \(q(w_i)\); endpoint terms in \eqref{eq:aft-u-deriv} when \(q\) is evaluated at \(u_i(\boldsymbol{\beta})\) &
\(v_i\) re-enters at every derivative order
\\
\bottomrule
\end{tabularx}
\end{table}

\section{Hessian block-inversion and numerical conditioning}\label{app:hessian-blocks}

Section~\ref{sec:delta-method} notes that the observed Hessian used for
delta-method inference is inverted block-wise, using separate blocks for
the regression coefficients \(\boldsymbol{\beta}\) and the basis parameters
\(\boldsymbol{\gamma}\). This block-wise inversion is a numerical
stabilisation strategy for cases in which the entire observed information
matrix is singular or nearly singular, so that direct inversion may be
unreliable \citep{nocedal2006numerical}. If one of the resulting blocks
requires additional numerical conditioning, a Cholesky--QR solve is used
in preference to a full matrix pseudoinverse, since the pseudoinverse can
mask, rather than resolve, near-singularity in a way that silently
degrades the reported standard errors. This conditioning choice is
internal to \texttt{vcov()} and \texttt{survfit.spbp} and does not require
any action from the user; the resulting instability warnings are described
in Section~\ref{sec:delta-method}.

\section{Monte Carlo numerical summaries}\label{app:mc-tables}

This appendix reports the summaries of the degree-sensitivity of LLPH\(\to\) BPPH.
Table~\ref{tab:bp-mcsim-abc} in the main text is the baseline Monte Carlo
grid. Figure~\ref{fig:degree-llph-bpph-plot} uses maximum likelihood only.
Within each replicate, the same simulated dataset is reused across degrees
\(m\in\{n^{0.2},\ldots,n^{0.8}\}\), so differences mainly reflect degree
rather than resampling noise.

\begin{figure}[!h]
\centering
\includegraphics[width=1\linewidth]{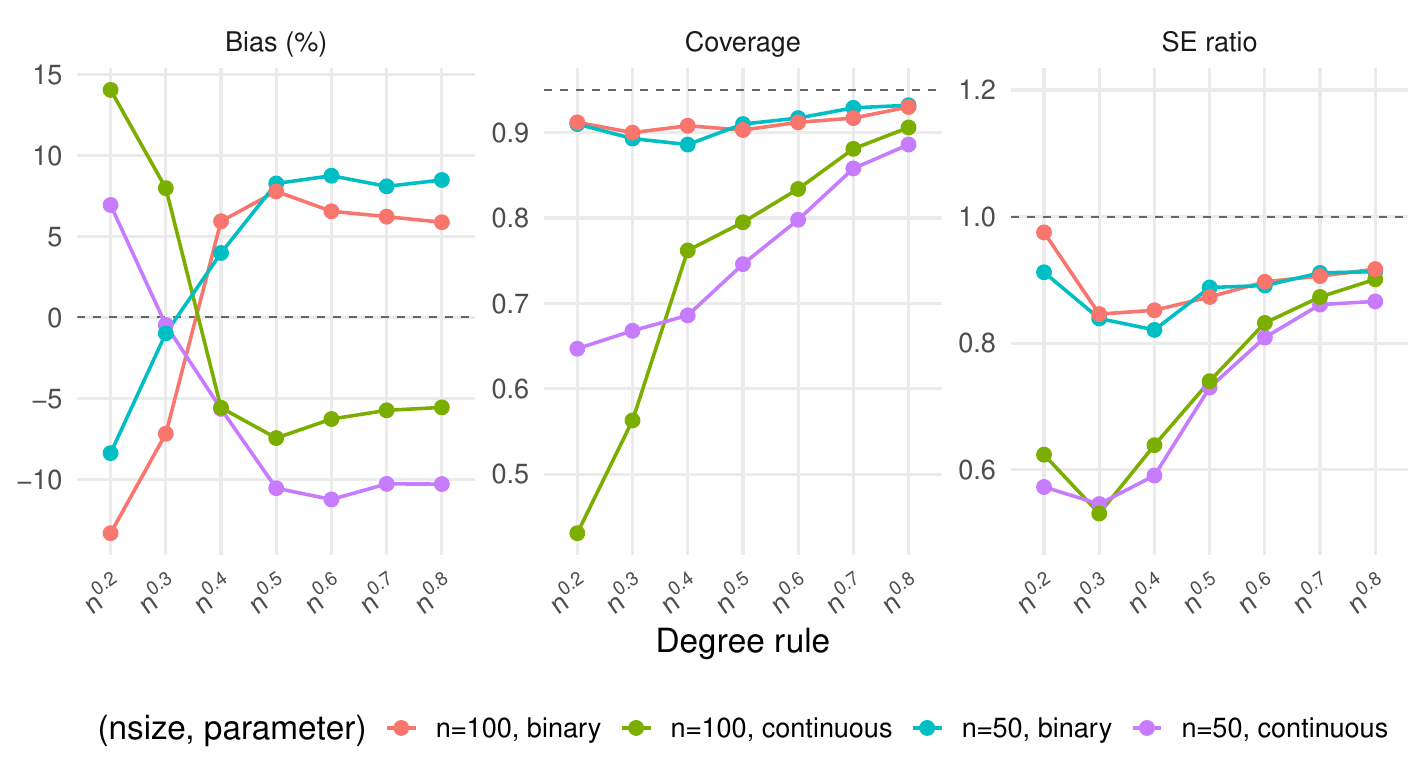}
\caption{Polynomial-degree sensitivity under log-logistic PH generation
with MLE BPPH. Degree rule on the horizontal axis; each line is one sample
size and parameter. Panels: 95\% Wald coverage, relative bias (\%), and SE
ratio. Full values in Table~\ref{tab:degree-llph-bpph}.}
\label{fig:degree-llph-bpph-plot}
\end{figure}

\begin{table}[!h]
\centering
\scriptsize
\setlength{\tabcolsep}{2.6pt}
\renewcommand{\arraystretch}{1.05}
\caption{\label{tab:degree-llph-bpph}
Degree sensitivity under LLPH/BPPH (\(R=1000\)). Coverage for 95\% Wald intervals; relative bias (\%); SE ratio. Degree rule \(m=n^p\) (see Figure~\ref{fig:degree-llph-bpph-plot}). Columns \(50\), \(100\), and \(200\) are sample sizes \(n\).}
\begin{tabular}{@{}ll*{12}{r}@{}}
\toprule
&&
\multicolumn{3}{c}{\(m\)} &
\multicolumn{3}{c}{Coverage (\%)} &
\multicolumn{3}{c}{Rel.\ bias (\%)} &
\multicolumn{3}{c}{SE ratio} \\
\cmidrule(lr){3-5}
\cmidrule(lr){6-8}
\cmidrule(lr){9-11}
\cmidrule(l){12-14}
Degree rule & Parameter & 50 & 100 & 200 & 50 & 100 & 200 & 50 & 100 & 200 & 50 & 100 & 200 \\
\midrule
\(n^{0.2}\) & Cont. (age) &  3 &  3 &  3 & 64.7 & 43.1 & 16.3 &  6.9 &  14.0 &  17.2 & 0.57 & 0.62 & 0.86 \\
\(n^{0.2}\) & Bin. (sex) &  3 &  3 &  3 & 91.0 & 91.2 & 85.9 & -8.4 & -13.3 & -17.3 & 0.91 & 0.97 & 1.19 \\
\(n^{0.3}\) & Cont. (age) &  4 &  4 &  5 & 66.8 & 56.3 & 60.5 & -0.5 &  8.0 &  6.5 & 0.55 & 0.53 & 0.57 \\
\(n^{0.3}\) & Bin. (sex) &  4 &  4 &  5 & 89.3 & 90.0 & 89.9 & -1.0 & -7.2 & -6.6 & 0.84 & 0.85 & 0.89 \\
\(n^{0.4}\) & Cont. (age) &  5 &  7 &  9 & 68.6 & 76.2 & 83.4 & -5.6 & -5.6 & -4.5 & 0.59 & 0.64 & 0.76 \\
\(n^{0.4}\) & Bin. (sex) &  5 &  7 &  9 & 88.6 & 90.8 & 93.2 &  4.0 &  5.9 &  4.3 & 0.82 & 0.85 & 0.93 \\
\(n^{0.5}\) & Cont. (age) &  8 & 10 & 15 & 74.6 & 79.5 & 88.9 & -10.5 & -7.4 & -4.2 & 0.73 & 0.74 & 0.91 \\
\(n^{0.5}\) & Bin. (sex) &  8 & 10 & 15 & 91.0 & 90.3 & 94.1 &  8.3 &  7.8 &  3.8 & 0.89 & 0.87 & 1.00 \\
\(n^{0.6}\) & Cont. (age) & 11 & 16 & 25 & 79.8 & 83.4 & 91.3 & -11.2 & -6.3 & -3.5 & 0.81 & 0.83 & 0.95 \\
\(n^{0.6}\) & Bin. (sex) & 11 & 16 & 25 & 91.7 & 91.2 & 94.7 &  8.7 &  6.6 &  3.0 & 0.89 & 0.90 & 1.02 \\
\(n^{0.7}\) & Cont. (age) & 16 & 26 & 41 & 85.8 & 88.1 & 92.6 & -10.3 & -5.7 & -3.5 & 0.86 & 0.87 & 0.96 \\
\(n^{0.7}\) & Bin. (sex) & 16 & 26 & 41 & 92.9 & 91.7 & 94.9 &  8.1 &  6.2 &  2.9 & 0.91 & 0.91 & 1.02 \\
\(n^{0.8}\) & Cont. (age) & 23 & 40 & 70 & 88.6 & 90.6 & 93.1 & -10.3 & -5.5 & -3.4 & 0.87 & 0.90 & 0.96 \\
\(n^{0.8}\) & Bin. (sex) & 23 & 40 & 70 & 93.2 & 93.0 & 94.5 &  8.5 &  5.9 &  2.9 & 0.91 & 0.92 & 1.01 \\
\bottomrule
\end{tabular}
\end{table}

A supplementary \(n=500\) grid (\(R=1000\), same LLPH\(\to\)BPPH design)
gives age coverage 7.2, 78.9, 88.8, 92.7, 92.4, 92.2, and 87.6\%
at \(m=n^{0.2},\ldots,n^{0.8}\) (\(m=4,7,13,23,42,78,145\)).
The package default \(m=\sqrt{n}=23\) is the peak; \(m=n^{0.4}=13\)
is the rule used in Table~\ref{tab:bp-mcsim-abc}.

\address{%
Renato Valladares Panaro\\
University Medical Center G\"{o}ttingen\\%
Department of Medical Statistics\\ Georg-August-Universit\"{a}t\\ 37073 G\"{o}ttingen, Germany\\
\url{https://rvpanaro.github.io/}\\%
\href{mailto:rvpanaro@gmail.com}{\nolinkurl{rvpanaro@gmail.com}}%
}

\address{%
Vin\'{i}cius Mayrink\\
Universidade Federal de Minas Gerais\\%
Departamento de Estat\'{i}stica, ICEx\\
Av.\ Ant\^{o}nio Carlos 6627, Pampulha\\
31270-901 Belo Horizonte, MG, Brazil\\
\url{http://www.est.ufmg.br/~vdinizm/}\\%
\href{mailto:vdinizm@gmail.com}{\nolinkurl{vdinizm@gmail.com}}%
}

\address{%
F\'{a}bio Demarqui\\
Universidade Federal de Minas Gerais\\%
Departamento de Estat\'{i}stica, ICEx\\
Av.\ Ant\^{o}nio Carlos 6627, Pampulha\\
31270-901 Belo Horizonte, MG, Brazil\\
\url{http://www.est.ufmg.br/~fndemarqui/}\\%
\href{mailto:fndemarqui@gmail.com}{\nolinkurl{fndemarqui@gmail.com}}%
}


\clearpage

\begin{thebibliography}{44}
\providecommand{\natexlab}[1]{#1}
\expandafter\ifx\csname url\endcsname\relax
  \def\url#1{\texttt{#1}}\fi
\expandafter\ifx\csname urlprefix\endcsname\relax\def\urlprefix{URL }\fi
\expandafter\ifx\csname href\endcsname\relax
  \def\href#1#2{#2} \def\path#1{#1}\fi

\bibitem[{R Core Team}(2025)]{R:2025}
{R Core Team}, R: A Language and Environment for Statistical Computing, R Foundation for Statistical Computing, Vienna, Austria (2025).
\newline\urlprefix\url{https://www.R-project.org/}

\bibitem[{Terry M. Therneau} and {Patricia M. Grambsch}(2000)]{survival:2000}
T.~M.~Therneau, P.~M.~Grambsch, Modeling Survival Data: Extending the {C}ox Model, Springer, New York, 2000.

\bibitem[Jackson(2016)]{flexsurv:2016}
C.~Jackson, {flexsurv}: A platform for parametric survival modeling in {R}, Journal of Statistical Software 70~(8) (2016) 1--33.
\newblock \href {https://doi.org/10.18637/jss.v070.i08} {\path{doi:10.18637/jss.v070.i08}}.

\bibitem[Scheike and Zhang(2011)]{timereg}
T.~H.~Scheike, M.-J.~Zhang, Analyzing competing risk data using the {R} {timereg} package, Journal of Statistical Software 38~(2) (2011) 1--15.
\newline\urlprefix\url{http://www.jstatsoft.org/v38/i02/}

\bibitem[Sjoberg and Baillie(2024)]{ggsurvfit:2024}
D.~D.~Sjoberg, M.~Baillie, ggsurvfit: Flexible Time-to-Event Figures, R package version 1.2.0 (2025).
\newline\urlprefix\url{https://CRAN.R-project.org/package=ggsurvfit}

\bibitem[Zhou et~al.(2021)Zhou, Hanson, and Zhang]{Zhou:2021}
H.~Zhou, T.~Hanson, J.~Zhang, {spBayesSurv}: Bayesian Modeling and Analysis of Spatially Correlated Survival Data, R package version 1.1.5 (2021).
\newline\urlprefix\url{https://CRAN.R-project.org/package=spBayesSurv}

\bibitem[Panaro(2026)]{spsurv:2026}
R.~Panaro, spsurv: Bernstein Polynomial Based Semiparametric Survival Analysis, R package version 1.1.0 (2026).
\newline\urlprefix\url{https://CRAN.R-project.org/package=spsurv}
\newblock \href {https://doi.org/10.32614/CRAN.package.spsurv} {\path{doi:10.32614/CRAN.package.spsurv}}.

\bibitem[Kalbfleisch and Prentice(2002)]{kalbfleisch2002statistical}
J.~D.~Kalbfleisch, R.~L.~Prentice, The Statistical Analysis of Failure Time Data, second Edition, Wiley Series in Probability and Statistics, Wiley, New York, 2002.
\newblock \href {https://doi.org/10.1002/9781118032985} {\path{doi:10.1002/9781118032985}}.

\bibitem[Tenbusch(1997)]{Tenbusch:1997}
A.~Tenbusch, Nonparametric curve estimation with Bernstein estimates, Metrika 45~(1) (1997) 1--30.
\newblock \href {https://doi.org/10.1007/BF02717090} {\path{doi:10.1007/BF02717090}}.

\bibitem[Chang et~al.(2007)Chang, Chien, Hsiung, Wen, and Wu]{Chang:2007}
I.-S.~Chang, L.-C.~Chien, C.~A.~Hsiung, C.-C.~Wen, Y.-J.~Wu, Shape restricted regression with random Bernstein polynomials, Institute of Mathematical Statistics Lecture Notes--Monograph Series 54 (2007) 187--202.
\newblock \href {https://doi.org/10.1214/074921707000000157} {\path{doi:10.1214/074921707000000157}}.

\bibitem[Vitale(1975)]{Vitale:1975}
R.~A.~Vitale, A Bernstein polynomial approach to density function estimation, in: M.~L.~Puri (Ed.), Statistical Inference and Related Topics, Academic Press, 1975, pp.~87--99.
\newblock \href {https://doi.org/10.1016/B978-0-12-568002-8.50011-2} {\path{doi:10.1016/B978-0-12-568002-8.50011-2}}.

\bibitem[Petrone(1999)]{Petrone:1999}
S.~Petrone, Bayesian density estimation using Bernstein polynomials, Canadian Journal of Statistics 27~(1) (1999) 105--126.
\newblock \href {https://doi.org/10.2307/3315494} {\path{doi:10.2307/3315494}}.

\bibitem[Babu et~al.(2002)Babu, Canty, and P.Chaubey]{Babu:2002}
G.~J.~Babu, A.~J.~Canty, Y.~P.~Chaubey, Application of Bernstein polynomials for smooth estimation of a distribution and density function, Journal of Statistical Planning and Inference 105~(2) (2002) 377--392.
\newblock \href {https://doi.org/10.1016/S0378-3758(01)00265-8} {\path{doi:10.1016/S0378-3758(01)00265-8}}.

\bibitem[Cox(1972)]{Cox:1972}
D.~R.~Cox, Regression models and life-tables, Journal of the Royal Statistical Society, Series B 34~(2) (1972) 187--202.
\newblock \href {https://doi.org/10.1111/j.2517-6161.1972.tb00899.x} {\path{doi:10.1111/j.2517-6161.1972.tb00899.x}}.

\bibitem[Chang et~al.(2005)Chang, Hsiung, Wu, and Yang]{Chang:2005}
I.-S.~Chang, C.~A.~Hsiung, Y.-J.~Wu, C.-C.~Yang, Bayesian survival analysis using Bernstein polynomials, Scandinavian Journal of Statistics 32~(3) (2005) 447--466.
\newblock \href {https://doi.org/10.1111/j.1467-9469.2005.00451.x} {\path{doi:10.1111/j.1467-9469.2005.00451.x}}.

\bibitem[Osman and Ghosh(2012)]{Osman:2012}
M.~Osman, S.~K.~Ghosh, Nonparametric regression models for right-censored data using Bernstein polynomials, Computational Statistics and Data Analysis 56~(3) (2012) 559--573.
\newblock \href {https://doi.org/10.1016/j.csda.2011.08.019} {\path{doi:10.1016/j.csda.2011.08.019}}.

\bibitem[McLain and Ghosh(2013)]{Mclain:2013}
A.~C.~McLain, S.~K.~Ghosh, Efficient sieve maximum likelihood estimation of time-transformation models, Journal of Statistical Theory and Practice 7~(2) (2013) 285--303.
\newblock \href {https://doi.org/10.1080/15598608.2013.772835} {\path{doi:10.1080/15598608.2013.772835}}.

\bibitem[Chen et~al.(2014)Chen, Hanson, and Zhang]{Chen:2014}
Y.~Chen, T.~Hanson, J.~Zhang, Accelerated hazards model based on parametric families generalized with Bernstein polynomials, Biometrics 70~(1) (2014) 192--201.
\newblock \href {https://doi.org/10.1111/biom.12104} {\path{doi:10.1111/biom.12104}}.

\bibitem[Zhou and Hanson(2018)]{Zhou:2018}
H.~Zhou, T.~Hanson, A unified framework for fitting Bayesian semiparametric models to arbitrarily censored survival data, including spatially referenced data, Journal of the American Statistical Association 113~(522) (2018) 571--581.
\newblock \href {https://doi.org/10.1080/01621459.2017.1356316} {\path{doi:10.1080/01621459.2017.1356316}}.

\bibitem[Klein and Moeschberger(1997)]{Klein:1997}
J.~P.~Klein, M.~L.~Moeschberger, Survival Analysis: Techniques for Censored and Truncated Data, Springer, New York, 1997.

\bibitem[Collett(2015)]{Collett2015}
D.~Collett, Modelling Survival Data in Medical Research, third Edition, Chapman and Hall/CRC, Boca Raton, 2015.

\bibitem[Bennett(1983)]{Bennett:1983}
S.~Bennett, Analysis of survival data by the proportional odds model, Statistics in Medicine 2~(2) (1983) 273--277.

\bibitem[Prentice(1973)]{Prentice:1973}
R.~L.~Prentice, Exponential survivals with censoring and explanatory variables, Biometrika 60~(2) (1973) 279--288.

\bibitem[Nocedal and Wright(2006)]{nocedal2006numerical}
J.~Nocedal, S.~J.~Wright, Numerical Optimization, second Edition, Springer, New York, 2006.

\bibitem[Gelman et~al.(2008)Gelman, Jakulin, Pittau, and Su]{gelman2008weakly}
A.~Gelman, A.~Jakulin, M.~G.~Pittau, Y.-S.~Su, A weakly informative default prior distribution for logistic and other regression models, The Annals of Applied Statistics 2~(4) (2008) 1360--1383.
\newblock \href {https://doi.org/10.1214/08-AOAS191} {\path{doi:10.1214/08-AOAS191}}.

\bibitem[Demarqui(2024)]{rsurv}
F.~Demarqui, rsurv: Random Generation of Survival Data, R package version 0.0.2 (2024).
\newline\urlprefix\url{https://CRAN.R-project.org/package=rsurv}

\bibitem[de~Mello~e Silva et~al.(2025)de~Mello~e Silva, Ghosh, and Mayrink]{de2025degree}
J.~F. de~Mello~e~Silva, S.~K.~Ghosh, V.~D.~Mayrink, Degree selection methods for curve estimation via Bernstein polynomials, Computational Statistics 40~(1) (2025) 1--26.

\bibitem[Kardaun(1983)]{Kardaun:1983}
O.~Kardaun, Statistical survival analysis of male larynx-cancer patients-a case study, Statistica Neerlandica 37~(3) (1983) 103--125.

\bibitem[Spiegelhalter et~al.(2002)Spiegelhalter, Best, Carlin, and Van Der~Linde]{spiegelhalter2002bayesian}
D.~J.~Spiegelhalter, N.~G.~Best, B.~P.~Carlin, A.~Van~Der~Linde, Bayesian measures of model complexity and fit, Journal of the Royal Statistical Society, Series B 64~(4) (2002) 583--639.

\bibitem[Vehtari et~al.(2017)Vehtari, Gelman, and Gabry]{Vehtari:2017}
A.~Vehtari, A.~Gelman, J.~Gabry, Practical Bayesian model evaluation using leave-one-out cross-validation and {WAIC}, Statistics and Computing 27~(5) (2017) 1413--1432.
\newblock \href {https://doi.org/10.1007/s11222-016-9696-4} {\path{doi:10.1007/s11222-016-9696-4}}.

\bibitem[Watanabe(2013)]{watanabe2013widely}
S.~Watanabe, A widely applicable Bayesian information criterion, Journal of Machine Learning Research 14~(27) (2013) 867--897.

\bibitem[Geisser and Eddy(1979)]{geisser1979predictive}
S.~Geisser, W.~F.~Eddy, A predictive approach to model selection, Journal of the American Statistical Association 74~(365) (1979) 153--160.

\bibitem[Ibrahim et~al.(2001)Ibrahim, Chen, and Sinha]{ibrahim2001bayesian}
J.~G.~Ibrahim, M.-H.~Chen, D.~Sinha, Bayesian Survival Analysis, Springer Series in Statistics, Springer, New York, 2001.
\newblock \href {https://doi.org/10.1007/978-1-4757-3447-8} {\path{doi:10.1007/978-1-4757-3447-8}}.

\bibitem[Pettitt(1984)]{Pettitt:1984}
A.~N.~Pettitt, Proportional odds models for survival data and estimates using ranks, Journal of the Royal Statistical Society, Series C (Applied Statistics) 33~(2) (1984) 169--175.
\newblock \href {https://doi.org/10.2307/2347443} {\path{doi:10.2307/2347443}}.

\bibitem[Therneau(2020)]{Therneau:2020}
T.~M.~Therneau, {survival}: Survival Analysis, R package version 3.2-7 (2020).
\newline\urlprefix\url{https://CRAN.R-project.org/package=survival}

\bibitem[Barron et~al.(1999)Barron, Birg\'{e}, and Massart]{barron1999risk}
A.~Barron, L.~Birg\'{e}, P.~Massart, Risk bounds for model selection via penalization, Probability Theory and Related Fields 113~(3) (1999) 301--413.
\newblock \href {https://doi.org/10.1007/s004400050210} {\path{doi:10.1007/s004400050210}}.

\bibitem[Bedrick et~al.(1996)Bedrick, Christensen, and Johnson]{bedrick1996priors}
E.~J.~Bedrick, R.~Christensen, W.~Johnson, A new perspective on priors for generalized linear models, Journal of the American Statistical Association 91~(436) (1996) 1450--1460.
\newblock \href {https://doi.org/10.1080/01621459.1996.10476713} {\path{doi:10.1080/01621459.1996.10476713}}.

\bibitem[Casella and Berger(2002)]{Casella:2002}
G.~Casella, R.~L.~Berger, Statistical Inference, second Edition, Duxbury, Pacific Grove, 2002.

\bibitem[Chen(2007)]{chen2007sieve}
X.~Chen, Large sample sieve estimation of semi-nonparametric models, in: J.~J.~Heckman, E.~E.~Leamer (Eds.), Handbook of Econometrics, Vol.~6B, Elsevier, 2007, pp.~5549--5632.
\newblock \href {https://doi.org/10.1016/S1573-4412(07)06076-X} {\path{doi:10.1016/S1573-4412(07)06076-X}}.

\bibitem[Farouki(2012)]{farouki2012bernstein}
R.~T.~Farouki, The Bernstein polynomial basis: A centennial retrospective, Computer Aided Geometric Design 29~(6) (2012) 379--419.

\bibitem[Farouki and Rajan(1987)]{Farouki:1987}
R.~T.~Farouki, V.~T.~Rajan, On the numerical condition of polynomials in Bernstein form, Computer Aided Geometric Design 4~(3) (1987) 191--216.
\newblock \href {https://doi.org/10.1016/0167-8396(87)90012-4} {\path{doi:10.1016/0167-8396(87)90012-4}}.

\bibitem[Lorentz(1953)]{Lorentz:1953}
G.~G.~Lorentz, Bernstein Polynomials, Mathematical Expositions, University of Toronto Press, Toronto, 1953.

\bibitem[Oehlert(1992)]{oehlert1992note}
G.~W.~Oehlert, A note on the delta method, The American Statistician 46~(1) (1992) 27--29.

\bibitem[Shen and Wong(1994)]{shen1994convergence}
X.~Shen, W.~H.~Wong, Convergence rate of sieve estimates, The Annals of Statistics 22~(2) (1994) 580--615.
\newblock \href {https://doi.org/10.1214/aos/1176325486} {\path{doi:10.1214/aos/1176325486}}.

\end{thebibliography}
\end{document}